\documentclass[oldversion]{aa}
\usepackage{graphicx}
\usepackage{times}
\usepackage{times,graphics}
\usepackage{lscape}
\usepackage{longtable}
\usepackage{epsfig} 
\usepackage{txfonts} 
\usepackage{natbib}
\usepackage{psfig}
\usepackage{subfig}
\bibpunct{(}{)}{;}{a}{}{,} 
\def\5{\footnotesize V\normalsize}
\def\4{\footnotesize IV\normalsize}
\def\3{\footnotesize III\normalsize}
\def\2{\footnotesize II\normalsize}
\def\1{\footnotesize I\normalsize}

\def\kms{$\mbox{km s}^{-1}$}

\def\fwhm{{\sc fwhm}}
\begin{document}

\title{Possible Evidence of Asymmetry in SN~2007rt, a Type IIn Supernova. }
   \author{C. Trundle \inst{1}, A. Pastorello \inst{1}, S. Benetti,
   \inst{2},  R. Kotak \inst{1}, S. Valenti \inst{1}, I. Agnoletto \inst{2}, F. Bufano \inst{2}, M. Dolci \inst{3}, N. Elias-Rosa \inst{4}, T. Greiner \inst{5}, D. Hunter \inst{1},  F.P. Keenan \inst{1}, V. Lorenzi\inst{6}, K. Maguire \inst{1}, S. Taubenberger \inst{7}} 

   \offprints{C.Trundle,~\email{c.trundle@qub.ac.uk.}}

   \authorrunning{Trundle et~al.}
   
   \titlerunning{SN~2007rt}
    
   \institute{Astronomy Research Centre, Department of Physics \& Astronomy,
   School of Mathematics \& Physics, The Queen's University of Belfast, Belfast, BT7 1NN, Northern Ireland
\and  INAF Osservatorio Astronomico di Padova, Vicolo dell'Osservatorio 5, 35122 Padova, Italy 
\and Spitzer Science Center, California Institute of Technology, 1200 E. California Blvd., Pasadena, CA 91125, USA
\and INAF Osservatorio Astronomico di Collurania, via M. Maggini, 64100 Teramo, Italy
\and SLOOH Telescopes, 176 West Morris Road, Washington CT, 06793, USA
\and Fundac\'ion Galileo Galilei-INAF, Telescopio Nazionale Galileo, 38700 Santa Cruz de la Palma, Tenerife, Spain
\and Max-Planck-Institut f\"{u}r Astrophysik, Karl-Schwarzschild-Str. 1, 85741 Garching bei M\"{u}nchen, Germany
}

   \date{}

   \abstract{An optical photometric and spectroscopic analysis of the slowly-evolving Type IIn SN~2007rt is presented, covering a duration of 481 days after discovery. Its earliest spectrum, taken approximately 100 days after the explosion epoch, indicates the presence of a dense circumstellar medium, with which the supernova ejecta is interacting. This is supported by the slowly-evolving light curve.  A notable feature in the spectrum of SN~2007rt is the presence of a broad He {\sc i} 5875 line, not usually detected in Type IIn supernovae. This may imply that the progenitor star has a high He/H ratio, having shed a significant portion of its hydrogen shell via mass-loss.  An intermediate resolution spectrum reveals a narrow H$_{\alpha}$ P-Cygni profile, the absorption component of which has a width of 128 \kms. This slow velocity suggests that the progenitor of SN~2007rt recently underwent mass-loss with wind speeds comparable to the lower limits of those detected in luminous blue variables. Asymmetries in the line profiles of H and He at early phases bears some resemblance to double-peaked features observed in a number of Ib/c spectra. These asymmetries may be indicative of an asymmetric or bipolar outflow or alternatively dust formation in the fast expanding ejecta. In addition, the late time spectrum, at over 240 days post-explosion, shows clear evidence for the presence of newly formed dust.}
 \maketitle  
%
\section{Introduction}  
Stars with masses greater than 7-8 M$_{\odot}$ are thought to end their lives as core-collapse supernovae (CCSNe). During their lifetimes massive stars undergo significant mass-loss, particularly when massive enough to pass through a luminous blue variable (LBV) or Wolf-Rayet (WR) phase. Not surprisingly, this mass-loss leaves behind circumstellar material (CSM) surrounding the star. 

Evidence for the presence of surrounding CSM has been detected in a number of  hydrogen rich SNe. The spectra of these objects do not show the broad P-Cygni profiles of the prototypical Type II supernovae. Instead, they are distinguished by their narrow H$_{\alpha}$ emission ($<$ 1000 \kms) on top of a broader emission profile.  This narrow feature is a signature for the presence of circumstellar matter previously shed by the progenitor star. Such objects led \citet{sch90} to identify a sub-class of objects amongst the hydrogen rich core-collapse supernovae, namely Type IIn. 

It is now generally believed that if a star undergoes significant mass-loss in its lifetime and subsequently evolves into a supernova, the fast moving ejecta from the explosion interacts with the discarded wind-material. This interaction is believed to cause a fast shock wave in the CSM and a reverse shock in the ejecta, with the shocked regions emitting high energy radiation \citep{chev94,chug94}. The intensity of this interaction and its effect on the supernova spectrum and light-curve is dependent on the density, composition and geometrical configuration of the CSM, and can provide an excellent trace of the mass-loss in the pre-explosion lifetime of the progenitor star. Type IIn SNe constitute a very heterogeneous group of objects showing a wide variation in the strengths of their emission lines and behaviour of their light curves.  Whilst the prototype IIn, SN~1988Z \citep{stat91,tur93,art99} is less luminous than Type Ia SNe, some of the most luminous SNe belong to this class \citep[viz. 2006gy, 2006tf; see][]{ofek07,smith07,smith08,ag09}. 

In addition to the characteristic narrow H features, type IIn spectra generally have strong blue continua, to which a single blackbody can not provide an adequate fit whilst simultaneously fitting the red part of the spectrum. \cite{smith08b} suggested that at late times the blue continuum present in the type IIn SN~2005ip was a result of the presence of a forest of high ionisation forbidden emission lines and they refer to this as a pseudo-continuum. This verified earlier speculation by \citeauthor{stat91} and  \citeauthor{tur93} that the source of the blue spectral region in SN~1988Z was a strong interaction with the surrounding circumstellar material. 

The presence of CSM provides a unique insight into the mass-loss history of the progenitor prior to core-collapse. However, there remains an element of ambiguity over the progenitors of type IIn supernovae. Recent work has indicated that some of these objects may be connected to luminous blue variables (LBVs) or at least have undergone LBV-like behaviour shortly before core-collapse \citep[see][]{k06,GY07,GY08,smith07,tru08,smith08,ag09}. Another group, the hybrid Type Ia/IIn, are thought to be a Type Ia disguised as a Type IIn, due to the strong narrow H$_{\alpha}$ emission in their spectra and the possible presence of the S {\sc ii} and Si {\sc ii} features typical of Type Ia \citep[viz. SN~2002ic; see][]{h03,ald06,k04}. However, this is largely under debate within the community \citep{b06,tru08}. There are also a number of so-called `transitional' objects, where the presence of  varying degrees of narrow hydrogen and helium lines in their spectra place them in a classification scheme between Type IIn and Type Ib/c objects \citep[such as SN~2005la and the type Ibn, SN~2006jc][]{pas07,fol07,pas08a,pas08b,smith08a}. The ambiguities surrounding Type IIn progenitors leads us to tread carefully whilst discussing this group of supernovae, and justifies an in-depth analysis of those in the class which are dissimilar to the groups prototype, SN~1988Z  \citep{stat91,tur93,art99}.

In this paper we will discuss the photometric and spectroscopic evolution of SN~2007rt, for more than 400 days post-discovery. SN~2007rt was discovered by \citet{li07} in UGC~6109, from unfiltered KAIT images on the 24$^{\rm th}$ November 2007. \citet{blond07}, as part of the CfA Supernova Survey, classified SN~2007rt as a type IIn supernova, 2-3 months past maximum, and claim the two best comparison spectra for this object are of the type IIn's SN~1998S and 1996L.  However from followup spectra we identified a broad helium feature, which is not detected in SN~1998S or many other type IIn SNe, and hence warrants further investigation. 


\section{Observations} 
\label{obs}
\begin{center}
\begin{table*}
\hspace{-30cm}\caption[Observation log]
{\label{obslog2} 
Log of photometric and spectroscopic observations of SN~2007rt. }
\begin{flushleft}
\centering
\begin{tabular}{lclcccc} \hline \hline
 DATE   &      Phase  & Telescope/Instr & \multicolumn{3}{c}{Spectroscopy} & Photometry \\
        &      (days)*             &                  & Grism & Wav. Range (\AA)& Res.(\AA)   &   \\         
\hline 
\\
28-11-2007 &   85 &  ~~~~~~~~(1)                &         &           &      & BVRC \\
02-12-2007 &   89 &  ~~~~~~~~(1)              &         &           &      & BVRC \\
05-12-2007 &   92 &  ~~~~~~~~(2)           &         &           &      & VC   \\
09-12-2007 &   96 &  ~~~~~~~~(3)       &         &           &      & C    \\
15-12-2007 & 102 & TNG/Dolores      & LRB+LRR & 3200-9600 & 17   & UBVRI (ph)\\
19-12-2007 & 106 & Loiano1.52/BFOSC & gm4+gm5 & 3500-9600 & 14   &  BVR (ph)\\
13-01-2008 & 131 & NOT/ALFOSC       & gm4     & 3300-8900 & 13.5 & UBVRI\\
04-02-2008 & 153 & WHT/ISIS         & R1200R  & 6000-6850 & 0.7  &      \\
11-02-2008 & 160 & TNG/Dolores      & LRB+LRR & 3200-9600 & 9    & UBVRI\\
29-03-2008 & 207 & Ekar1.82/AFOSC   &	     &	         &      & BVRI \\
30-03-2008 & 208 & NOT/ALFOSC       & gm4     & 3300-8900 & 16.5 & UBVRI\\
29-04-2008 & 238 & Loiano1.52/BFOSC & gm4+gm5 & 3600-8700 & 13   &  \\
01-05-2008 & 240 & CAHA/CAFOS       & b200    & 3300-8900 & 13   & BVRI \\
03-06-2008 & 273 & CAHA/CAFOS       & b200    & 3300-8900 & 13   & UBVRI\\
07-07-2008 & 307 & CAHA/CAFOS       &	     &	         &      & UBVRI (ph)\\
05-12-2008 & 458 & NOT/ALFOSC       &        &          &      & R \\
22-12-2008 & 475 & TNG/Dolores        & LRR & 4950-9600& 13  & BVRI  \\
13-01-2009 & 497 & TNG/Dolores        & LRB & 3400-7750&  15 & RI  \\
20-01-2009 & 504 & TNG/Dolores        &          &                     &       & R  \\
22-01-2009 & 506 & TNG/Dolores        & LRB &3500-7900 & 12 &   \\
23-01-2009 & 507 & TNG/Dolores        & LRB+LRR & 3400-9500 & 12 & UBVR \\
19-03-2009 & 562 & CAHA/CAFOS     &                     &                       &       & BVRI\\
\hline
\multicolumn{7}{l}{(1)0.35m f/11 Schmidt-Cassegrain telescope+Kodak KAF-3200ME CCD, SLOOH, Teide Observatory,}\\  
\multicolumn{7}{l}{Canary Islands, Spain; (2)0.25m f/3.8 Takahashi Epsilon telescope E250+SBIG ST-8XE CCD - Remote}\\  
\multicolumn{7}{l}{Astronomical Society Observatory, New Mexico, USA; (3)0.37m f14 Cassegrain Iowa Robotic Telescope+}\\ 
\multicolumn{7}{l}{FLI PL-09000 CCD, Winer Observatory, Arizona $\&$ University of Iowa, USA}\\ 
\multicolumn{7}{l}{* Phase refers to the number of days post-exposion, assuming a discovery date of 81 days post-explosion.}\\

\end{tabular}
\end{flushleft}
\end{table*}
\end{center}

\begin{center}
\begin{table*}
\hspace{-30cm}\caption[Observation log]
{\label{photval} 
Optical photometry of SN~2007rt.}
\begin{flushleft}
\centering
\begin{tabular}{lccccccccccc} \hline \hline
 DATE   &      Phase(days)  & U & U$_{err}$ & B & B$_{err}$ & V & V$_{err}$ & R & R$_{err}$ & I & I$_{err}$ \\
\hline 
\\
28-11-2007 & 85 &                &            & 17.600 & 0.388 & 17.177 & 0.224 & 16.862 & 0.297 &        &       \\ 
02-12-2007 & 89 &                &            & 17.624 & 0.172 & 17.263 & 0.186 & 16.900 & 0.222 &        &       \\
05-12-2007 & 92 &                &            &        &       & 17.285 & 0.090 &        &       &        &       \\
09-12-2007 & 96 &                &            &        &       &        &       & 16.935 & 0.273 &        &       \\
15-12-2007 &102 & 17.191 & 0.027 & 17.670 & 0.019 & 17.327 & 0.025 & 16.977 & 0.021 & 16.803 & 0.033 \\
19-12-2007 &106 &               &            & 17.648 & 0.046 & 17.334 & 0.023 & 16.943 & 0.098 & 16.744 & 0.150 \\ 
13-01-2008 &131& 17.541  & 0.026 & 17.885 & 0.018 & 17.490 & 0.021 & 17.057 & 0.021 & 17.042 & 0.026 \\
11-02-2008 &160& 17.623  & 0.039 & 17.968 & 0.020 & 17.602 & 0.023 & 17.077 & 0.022 & 17.069 & 0.030 \\
29-03-2008 &207&                &            & 18.166 & 0.054 & 17.882 & 0.090 & 17.171 & 0.066 & 17.321 & 0.070 \\
30-03-2008 &208 & 17.808 & 0.041 & 18.160 & 0.039 & 17.881 & 0.097 & 17.176 & 0.028 & 17.329 & 0.118 \\
01-05-2008 &240 &               &            & 18.289 & 0.028 & 18.001 & 0.045 & 17.339 & 0.047 & 17.331 & 0.059\\
03-06-2008 &273 & 18.059 & 0.073 & 18.335 & 0.034 & 18.200 & 0.047 & 17.516 & 0.061 & 17.531 & 0.101 \\
07-07-2008 &307 & 18.206 & 0.051 & 18.601 & 0.049 & 18.433 & 0.082 & 17.618 & 0.059 & 17.580 & 0.087 \\
05-12-2008 &458 &               &             &              &             &               &           & 18.565 & 0.070 &               &  \\
22-12-2008 &475 &               &             & 20.175 & 0.102 & 19.948  & 0.082& 18.622 & 0.043 & 18.932 &   0.060\\
13-01-2009 &497 &               &             &              &             &               &             & 18.881 & 0.090 & 19.088 &0.258 \\
20-01-2009 &506 &               &             &              &             &               &             & 18.866 & 0.148 &              & \\
23-01-2009 &507 & 19.675  &  0.071&  20.543& 0.079 & 20.250  & 0.149  & 18.884 & 0. 096&              & \\
19-03-2009 &562 &               &             &  21.179& 0.103 & 21.012 & 0.095 & 19.610 &0.071&19.600 &0.055\\
\hline
\multicolumn{12}{l}{The phase is given relative to the explosion date,  believed to be 81 days prior to the discovery date on 24$^{th}$ November 2007.}\\
\end{tabular}
\end{flushleft}
\end{table*}
\end{center}

Photometric and spectroscopic data of SN~2007rt were collected from November 2007 to March 2009. The details of these observations are logged in Table~\ref{obslog2} and are outlined below. 
\subsection{Photometry} 
\label{obsphot}
\begin{figure} 
\epsfig{file=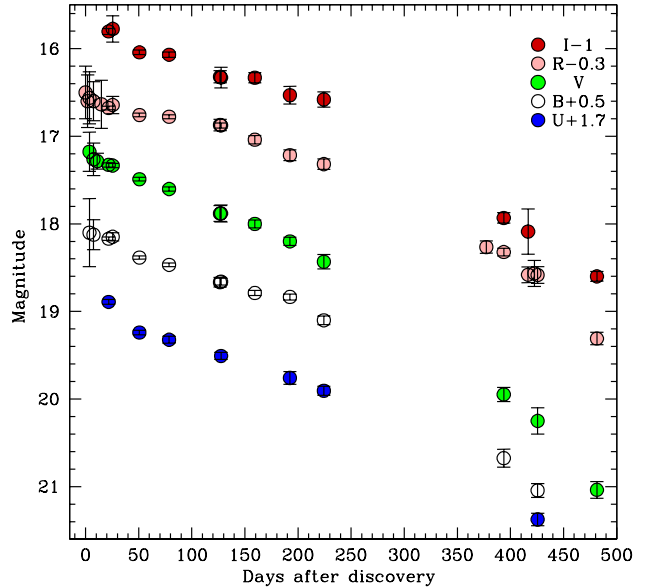, width=90mm ,angle=0} 
\vspace{-8mm}\caption[]{UBVRI light curves of SN~2007rt up to 481 days after discovery. From top to bottom the points are I, R, V, B, and U with offsets of -1, -0.3, 0, +0.5 and +1.7 mag. respectively. The discovery magnitudes reported by \citet{li07} are also included here. } 
\label{phot1} 
 \end{figure}

Our collaboration obtained optical photometry of SN~2007rt with the Telescopio Nazionale Galileo (TNG) and Nordic Optical Telescope (NOT) in La Palma (Canary Islands, Spain), the 1.82m Copernico telescope of the Asiago Observatory (Italy), the 1.52m telescope of the Loiano Observatory (Italy), and the 2.2m Calar Alto telescope (Spain).  In addition four data points provided by amateur astronomers were used. In total this gives a coverage from 4 to 481 days after discovery (see Fig.~\ref{phot1}). 
The images were trimmed, de-biased and flatfielded. Since template images of the host galaxy were not available, the SN magnitudes were measured using a point spread function (PSF) fitting technique in the Image Reduction and Analysis Facility (IRAF)\footnote{IRAF is distributed by the National Optical Astronomy Observatories, which are operated by the Association of Universities for Research in Astronomy, Inc., under cooperative agreement with the National Science Foundation.}. Zero-points were defined making use of standard Landolt fields observed on the same night as the SN. The magnitudes of SN~2007rt where then calibrated relative to the average magnitudes of a sequence of stars in the SN field obtained during three photometric nights (highlighted by (ph) in column 7 of Table~\ref{obslog2}). Unfiltered magnitudes obtained from amateur images were rescaled to the V or R band magnitudes depending on the wavelength position of the maximum of the quantum-efficiency curve of the detectors used \citep[see also the discussion in][]{pas08c}. These unfiltered magnitudes are denoted by C in Table~\ref{obslog2}. The calibrated SN magnitudes are presented in Table~\ref{photval} and Fig.~\ref{phot1}. 
\subsection{Spectroscopic Data}
\label{obsspec}
An intermediate resolution ($\sim$30 \kms) optical spectrum of SN~2007rt was obtained using the William Herschel telescope (WHT) on La Palma, as well as low resolution ($\sim$200 \kms) spectra from the telescopes listed above for photometry. These provided spectral coverage over a 426 day period from discovery on 24$^{\rm th}$ November 2007. Details of the epoch, wavelength coverage and resolution of these spectra are presented in Table~\ref{obslog2}. The spectra were reduced using standard spectral reduction procedures in IRAF.   They were wavelength and flux calibrated using arc lamps and spectrophotometric standards observed on the same night. The wavelength calibrations were verified using the narrow sky lines. Absolute flux calibrations were then made with photometry taken on the same night. Prior to flux calibration, it was necessary to apply a second order correction to the ALFOSC/NOT spectra, due to contamination in the red part of the spectra from blue light of the second order. This was accomplished using the procedure outlined by \cite{stan07}. 

The spectroscopic data have been corrected for the redshift of the host galaxy, UGC 06109 (z=0.022365 $\pm$ 0.00080)  as published in the updated Zwicky catalog \citep{fal99}. The blue continuum plus absence of Na {\sc i} D lines in the spectra of SN~2007rt suggest that there is a negligible effect due to extinction on the observed spectral energy distribution. Thus we have only corrected the spectra by the small Galactic extinction contribution as suggested by \citet{sch98}. Along the line of sight of SN~2007rt the Galactic extinction is E(B-V) = 0.02. 
%
\section{Light Curve evolution}
\label{photanalysis}
\begin{figure*} 
\begin{center}
\epsfig{file=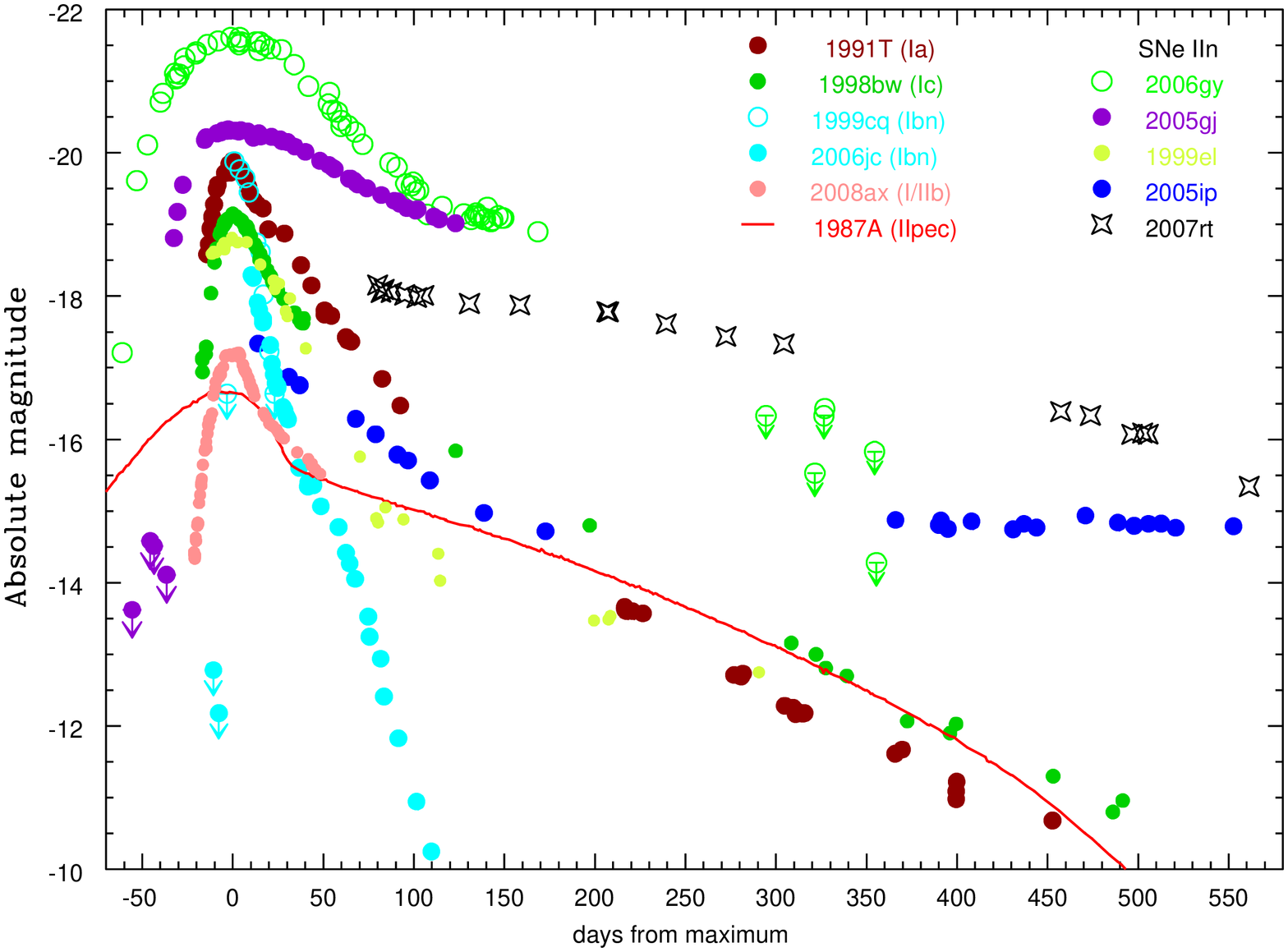,height=130 mm,angle=-90 }
\vspace{-8mm}\caption[]{Absolute R band magnitudes of SN~2007rt compared to a range
of SN types. Reference of the SNe magnitudes are: SN~1987A - \citet[and references therein]{whit89}, SN~1991T     - \citet{sch94,cap97,sparks99}, SN~1998bw - \citet{gal98,mck99,sol00,pat01}, SN~1999cq - \citet{mat00}, SN~2006jc - \citet{pas07,fol07},  SN~ 2008ax - \citet{pas08c}, and the Type IIn's  SN~1999el - \citet{dicarlo02}, SN~2005gj - \citet{ald06,p07}, SN~ 2005ip - \citet{smith08b}, SN~2006gy - \citet{smith07,ofek07,ag09,kaw09}.}
\label{phot2}  
\end{center}
\end{figure*}

Fig.~\ref{phot1} shows the UBVRI light curves of SN~2007rt. Over the first 130 days after discovery, the light curve  of SN~2007rt evolves slowly (at a rate of 0.003 mag d$^{-1}$)  and it is clear the supernova has not be caught at maximum (see Sect.~\ref{specevol}).  Following this there is a gradual decline in the light curve, which steepens at late phases (0.01 mag d$^{-1}$ from 458-562 days post-explosion). It was reported by \citet{li07} that on a KAIT image taken almost seven months prior to discovery, on 8$^{\rm th}$ May 2007, nothing was detected at the position of the supernova. Since this supernova has been discovered quite late on in its evolution,  as suggested by the non-detection of a maximum in the light curve, it is difficult to determine the explosion epoch. However we obtained an estimate of the explosion date of SN~2007rt indirectly using the spectra and light-curve of an interacting type IIn SN, SN~2005ip. 

\subsection{Age of SN~2007rt}
\label{age}
\begin{figure*}
\begin{center}
\epsfig{file=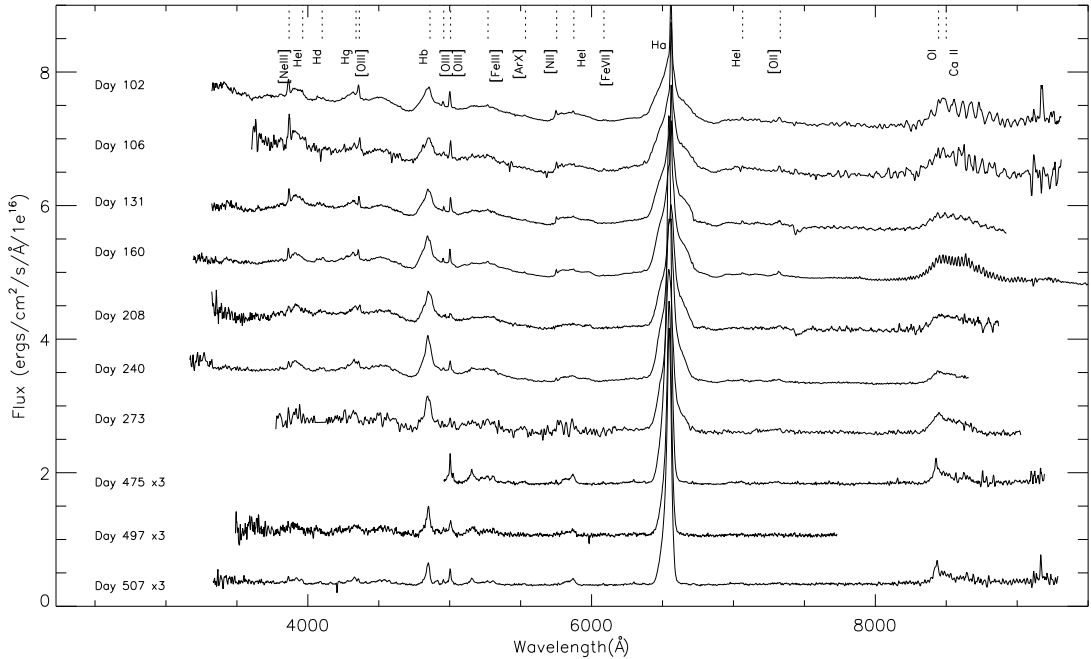, height=155 mm , angle=90}
\caption[]{Dereddened optical spectra of SN~2007rt from 102 to 507 days after its adopted explosion date. Flux for each spectrum is arbitrarily offset by 0.75 for clarity, and the spectra from days 475, 497, \& 507 are scaled by a factor of 3 in flux.}
\label{optspec} 
\end{center}
\end{figure*}

The light-curve of SN~2005ip declined rapidly after discovery, suggesting SN~2005ip was discovered close to maximum \citep{smith08b}.  Many core collapse SNe, for which the rise time to peak was observed, have rise times of 3 weeks or less. Few SNe Type IIns have been detected during this rise phase, and those that have, give rise times of 20-50 days (see SN~2006gy and SN~2005gj in Fig.~\ref{phot2}). The fact that type IIns are rarely detected before maximum, suggests that the rise time is quite fast. Based on this, it is reasonable to assume that SN~2005ip was detected within a couple of weeks from explosion. Hence we adopt an explosion date for SN~2005ip of JD=2453673, which is 7 days prior to its detection (Fig.~\ref{phot2}). This is consistent with the findings of \citeauthor{smith08b}, however we cannot ignore that there is a large degree of uncertainty in this. The assumption of a short rise time of SN2005ip is also supported by the rapid change in the spectra after discovery. The first spectrum shows an almost featureless and very blue spectrum  \citep{smith08b} and over the following days the broad features and continuum evolve rapidly (from unpublished spectra, Trundle et al. in prep).

At 95 days after its discovery, the spectrum of SN~2005ip best fits the broad features and slope of our first SN~2007rt spectrum (see Sect.~\ref{ccsne}).  Assuming the similarity of the spectra of SN~2005ip and SN~2007rt indicates that these supernovae are of similar age, the age estimate of the first spectral epoch of  SN~2007rt can be refined to approximately 102 $\pm$ 40 days after explosion, and thus the explosion date is set to JD=2454349 $\pm$ 40. Before adopting this age, for the following discussions, we will attempt to qualify our assumption that these two objects are close to or of a similar age at these epochs. The spectra of SN~2007rt at discovery and SN~2005ip at 95 days are heavily dominated by the CSM interaction. So although there is evidence for spectral similarity the properties of the CSM may mask the true epoch. However a very young age can be ruled out for SN~2007rt. Firstly, the first spectrum was obtained 21 days post-discovery and hence SN~2007rt must be at least 21 days old. Secondly, in an early epoch ($<$40 days) Type IIn,  a strong blue continuum would be expected which evolves rapidly along with the broad features (viz. SN~2005ip, SN~2005gj; \cite{smith08b,ald06}). However the blue continuum and broad features in SN~2007rt, do not evolve rapidly over a 60 day period between our first and fifth spectrum  (see further discussion on continuum Sect.~\ref{specevol}). Similarly a large amount of CSM, prolonging the interaction duration of  SN~2007rt could lead to an underestimate of its age. In spite of this, uncertainties of a few weeks in our estimates will not substantially alter our findings. From this point forward we adopt the explosion date of JD=2454349 $\pm$ 40, implying the observations of SN~2007rt were taken between 85 to 562 days post-explosion. \footnote{All dates in the rest of this article refer to the adopted explosion epoch, unless otherwise stated.}

The light curve of SN~2007rt is  compared to those of a range of SNe types in Fig.~\ref{phot2}. At approximately 100 days past maximum, SN~2007rt is more luminous than the fast-declining type IIn  SN~1999el, but fainter than some of the most luminous type IIn objects, SN~2006gy and SN~2005gj. Despite the similarity in the spectra of SN~2007rt and SN~2005ip at early times, there is a significant difference in the absolute luminosity of their light curves. This may be caused by differing densities and distributions of circumstellar matter surrounding the two SNe. The early time light-curve evolution of SN~2007rt is very slow, declining by 0.003 mag d$^{-1}$. This is considerably slower than SN~2005ip which has a decline rate for a similar period of 0.015 mag d$^{-1}$\citep{smith08b}. 

\section{Spectral evolution}
\label{specevol}
\begin{figure*}
\begin{center}
\epsfig{file=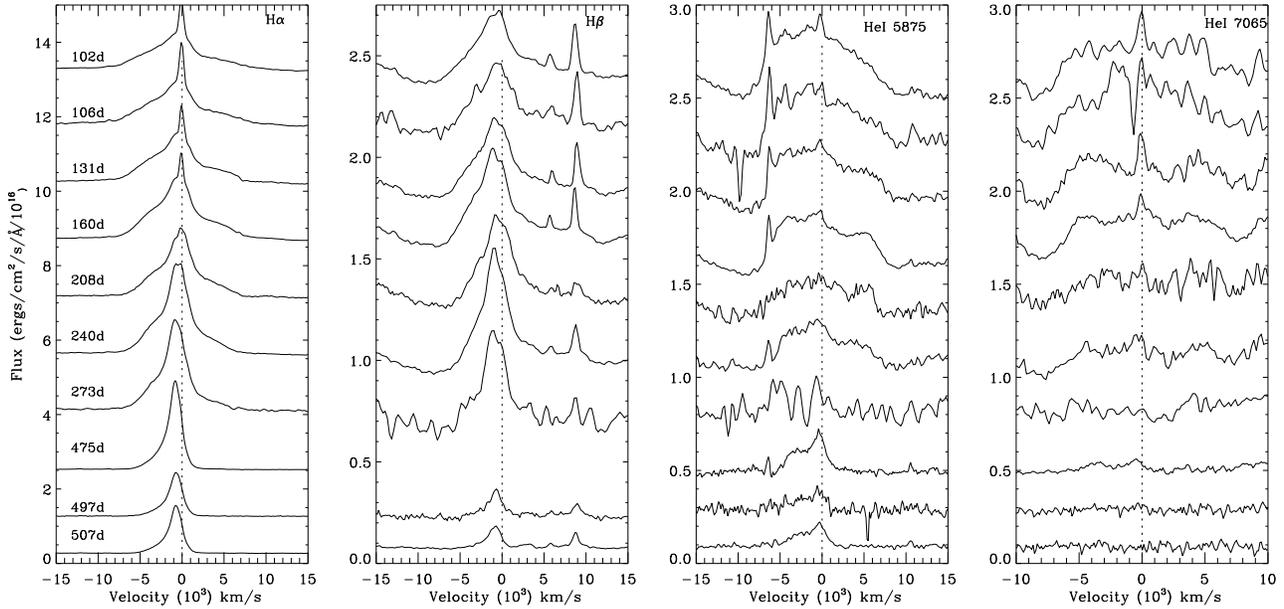, height=170 mm,  angle=90}
\caption[]{Observed H$_{\alpha}$, H$_{\beta}$, He {\sc i} 5875 \AA\ and He {\sc i} 7065 \AA~profiles in SN~2007rt from 102 to 507 days since the explosion date. The flux of each spectrum is arbitrarily offset  for clarity. Note the blue-shift in the later epochs of H$_{\alpha}$ and the asymmetry in its red wing. In the last two panels, the flux has been magnified by a factor of 4 for an easier view of the weak narrow helium features. Note under magnification there appears to be a weak broad helium component at 7065 \AA. }
\label{ha}
\end{center} 
\end{figure*}
The spectra of SN~2007rt, displayed in Fig.~\ref{optspec}, have been corrected for a Galactic reddening of E(B-V) = 0.02 mag as discussed in Sect.~\ref{obsspec}. There is a blue excess present in the spectra in the early epochs, which flattens after more than 300 days post-explosion. We fit a blackbody to the spectrum of SN~2007rt, and obtain a temperature of $\sim$7500 K for the first epoch. This fit in itself was poor and as the SN evolved it became increasingly difficult to simultaneously reconcile the flat red spectral region and the blue excess with a  single blackbody. \cite{smith08b} showed that, in the case of SN~2005ip, spectra taken some 200 days after maximum revealed a blue pseudo-continuum composed of a forest of narrow emission lines, rather than a real thermal continuum with a blackbody like behaviour. This indicated that for SN~2005ip the blue pseudo-continuum at late times arose from the interaction with circumstellar material, and was not a real thermal continuum. This had previously been speculated by \citet{stat91} and \citet{tur93} for the case of SN~1988Z. The low velocity of the CSM in SN~2005ip allowed the detection of the separate components of the emission line spectrum in the blue which formed the pseudo-continuum at late-stages.  It  therefore may be possible that the temperature derived by fitting a blackbody to the spectrum of SN~2007rt is not meaningful. This will be discussed further in Sect.~\ref{ccsne}.\footnote{From this point forward the term continuum refers to the overall shape of the spectrum, while `thermal continuum' and `pseudo-continuum'  are used to distinguish between a thermal blackbody like continuum and an unresolved forest of emission lines.}

Prominent features in the spectra of SN~2007rt are the narrow and broad H$_{\alpha}$ features, the intermediate velocity H$_{\beta}$ line and broad Ca {\sc ii} infrared triplet. There are also a number of narrow emission features associated with He and heavier metals in the spectra.  However, possibly the most peculiar spectral feature is the broad, relatively flat topped He {\sc i} 5875 \AA\ emission line.  In Fig.~\ref{optspec} these main features are identified. At all phases the H and He lines are asymmetric. At late times the H$_{\alpha}$ and He {\sc i }5875 \AA\ lines have a pronounced blueshift, and a significant lack of redshifted emission in the broad component. There is also a hint of a narrow component of He {\sc i} 5875 and 7065 \AA\  in the early epochs (see Fig.~\ref{ha}), suggesting there is a certain amount of helium in the CSM. The narrow He {\sc i} 5875 \AA\ line diminishes from the first spectra (102 days post-explosion), and disappears by day 208. In the later epochs, from day 475, an intermediate velocity component appears. This suggests that there is interaction between the ejecta and the He in the CSM. In addition, a strong permitted O {\sc i} 8446 \AA\ line develops from day 273,  which has velocities comparable to the intermediate H component.

To determine the properties of the spectral features we have fit Gaussians to their profiles. Each Gaussian was parameterized by three quantities, the central wavelength, full-width-half-maximum (\fwhm) and peak height. These parameters were freely fit by the fitting routine. In a small number of cases, the central wavelengths of the narrow lines were fixed, as the low intensity of the line relative to the continuum prohibited the detection of its peak. The resulting wavelengths, velocities and intensities of the lines are presented in Tables~\ref{hafittbl} - \ref{narrow}.  All \fwhm~given in the aforementioned tables and the discussions below are corrected for the instrumental resolution. In interacting supernovae the flux calibration is subject to large uncertainties due to the presence of strong emission lines. Errors on the measured intensities of the lines are therefore on the order of 20\%, whereas the {\sc fwhm} uncertainties are 10\%. The narrow lines which were detected and fitted in the spectrum had \fwhm~velocities that were not fully resolved by the low resolution spectra (these are discussed further in Sect.~\ref{narrowdis}). As a guide to the properties of these lines, their parameters as determined from the TNG spectrum taken on the 11$^{\rm th}$ February are presented in Table~\ref{narrow}. In addition a  number of the lines were resolved in the intermediate resolution WHT spectrum taken on the 4$^{\rm th}$ February, and these are also presented in the aforementioned table.

In the case of H$_{\alpha}$ and He {\sc i} 5875 \AA, multiple Gaussian fits were employed (see Sect.~\ref{haevol}) due to the asymmetry of the observed profiles. As discussed in more detail below, H$_{\alpha}$ was first fit with a broad, intermediate and narrow Gaussian whilst the He  {\sc i} 5875 \AA\ line was fit with a broad and narrow component for the early epochs, and a broad and intermediate component for the latest epochs. The parameters of the broad and intermediate components from these fits are shown in Table~\ref{hafittbl}. However these initial fits were poor, and hence we refit the profiles with up to four Gaussians; an intermediate component, two broad components, and where appropriate a narrow component. The parameters of the broad and intermediate components of the latter fits are presented in Table~\ref{hafittbl2}. 

\begin{center}
\begin{table}
\caption[Decomposition fits to H$_\alpha$]
 {\label{hafittbl} Initial parameters for the intermediate and broad Gaussian fits to spectral lines in SN~2007rt. }
\vspace{-4mm}
\begin{flushleft}
\centering
\begin{tabular}{llccc} \hline \hline
  Line                &   Day             &      $\lambda$                              & FWHM     & Intensity\\
                       &       &  \AA   &  \kms      & 10$^{-15}$erg.s$^{-1}$~cm$^{2}$\\ 
\hline 
He {\sc i} 3926  & 102 & 3919.12 &  7340  &  13.43  \\  
broad                 & 106 & 3919.36 &  6482   &  16.95  \\ 
                            & 131 & 3924.15 &  6463   &  11.85  \\ 
                            & 160 & 3918.89 &  6004   &    8.63   \\
                            & 208  & 3925.70 & 4850  &    6.97   \\ 
                            & 240  & 3921.51 &  4948 &    7.76  \\ 
                            & 273  & 3930.00 &  4155 &    6.53  \\ 
\hline
H$_{\gamma}$ 4340& 102  & 4318.15 &  5808   & 9.39  \\  
 Inter.                          & 106  & 4321.20  &  5114  & 6.31  \\ 
                                     & 131  & 4316.22 &  4851   & 6.99  \\ 
                                     & 160  & 4317.16  &  5683  & 7.37  \\
                                     & 208  & 4344.80  &  4649  & 6.74   \\ 
                                     & 240  & 4330.00  &  4743  &6.76   \\ 
                                     & 497  & 4337.32 & 3992   & 1.16 \\
\hline
H$_{\beta}$ 4861 & 102   & 4847.43 &  3850&  15.31  \\  
 Inter.                       & 106  & 4852.75  &  4151  &  16.80  \\ 
                                 & 131  & 4851.27 &  4046  &  19.64 \\ 
                                 & 160  & 484714  &  4099   &  24.52 \\
                                 & 208  & 4856.45 &  3638   &  20.63 \\ 
                                 & 240  & 4852.49  &  3258   &  24.78  \\ 
                                 & 273   & 4852.13  &  2678  & 16.04  \\ 
                                 & 497  & 4851.03& 1673  & 2.98\\
                                 & 507 & 4849.14 & 1614 & 2.35 \\
\hline
He {\sc i} 5875  & 102   & 5860.11 & 10279  &  24.11 \\  
 broad                 & 106   & 5872.61 &  9640    &  20.77  \\ 
                            & 131   & 5870.01 &  9729   &  19.05 \\ 
                            & 160   & 5869.47 & 10146   &  17.89  \\
                            & 208   & 5872.93 &  9834    &  11.56  \\ 
		         & 240   & 5867.50 &  8080    &   11.07   \\ 
		         & 475 & 5837.26 & 4466 & 1.74 \\		         
		         & 497  & 5848.50 & 4538 & 1.77 \\
		         &507 & 5835.03 & 4057 & 1.58 \\
\hline
He {\sc i} 5875  & 475   & 5872.62 & 1072  &  0.63\\  
Inter.*                 & 507  & 5872.61 &   1318   &  0.52 \\ 
		         
\hline
H$_{\alpha}$ 6563& 102   & 6543.71 & 10136   & 121.68  \\  
broad                        & 106  & 6546.19& 10113    & 145.75  \\ 
                                   & 131  & 6545.41&  8793     &  145.94 \\ 
                                   & 153  & 6543.88 &  8244     &    139.41 \\
                                   & 160  & 6550.00 &  8489     &   179.92 \\
                                   & 208 & 6553.00  &  7285    &    149.68  \\ 
                                   &240  & 6548.34   &  6835    &   171.70  \\ 
                                   & 273  & 6541.33  &  6047    &    129.30  \\ 
			       & 475 & 6529.07 & 3645 & 42.87 \\
                                   & 497  & 6530.32 &  3593    & 26.61  \\
                                   & 507 & 6530.72 & 3495 & 28.93 \\

\hline
H$_{\alpha}$ 6563& 102   & 6548.29 &  2905   &  24.11 \\  
Inter.                          & 106  & 6552.01 &  2687   & 27.02 \\ 
                                   & 131 & 6555.53 &  2179    &  24.30 \\ 
                                   & 153 & 6555.35 &  2195    & 29.38 \\
                                   & 160  & 6554.89 &  2212   &  33. 85 \\
                                   & 208  & 6562.25  &  1734  &  33.40  \\ 
                                   & 240  & 6557.58 &  1824   &  53.87  \\ 
                                   & 273  & 6553.28&  1812    &  53.08 \\
                                   & 475 & 6549.65 & 1353 & 40.20 \\   
                                   & 497  & 6551.11 & 1357 & 26.40  \\ 
                                   & 507 &6550.96 & 1335 & 27.72 \\
\hline
O {\sc i} 8446         & 273 & 8441.22 & 2980 & 12.59 \\
 inter.                       & 497 & 8433.39 & 1399 & 3.15 \\
                                 & 507&8434.35 & 1244 & 2.45\\
\hline
Ca {\sc ii} & 102 & 8550.00 & 12982   & 118.48 \\  
 broad      & 106  & 8540.47& 11541  & 101.92 \\ 
                 & 131 & 8553.22 & 11651  & 72.96  \\ 
                 & 160  & 8552.43 & 12486  & 117.55  \\
                 &273  & 8544.40&  8750  &  44.89 \\
                 &475 & 8535.883 & 8570 & 7.32 \\
\hline
\multicolumn{5}{l}{* An intermediate He {\sc i} 5875 \AA\ line was detected at 475 \& 507 days.  }\\
\multicolumn{5}{l}{Prior to that only weak narrow and broad components were present.}\\
 \end{tabular}
\end{flushleft}
\end{table}
\end{center}

\subsection{Narrow H$_{\alpha}$ P-Cygni profile}
\label{narrowha}
Fig.~\ref{ha} shows the time-series of the H$_{\alpha}$ line. From day 102 to day 208 there is a narrow emission component present which decreases in strength over time. The luminosity of this narrow component decreases by more than a factor of 2 between our first epoch of data and that taken 58 days later. It has disappeared completely 138 days after our first spectrum. This feature is not fully resolved by the low resolution spectra which makes up most of our dataset. However, we can place an upper limit on the \fwhm~velocity of $\leq$ 200 \kms (see Table~\ref{narrow}).  

\begin{figure}
\epsfig{file=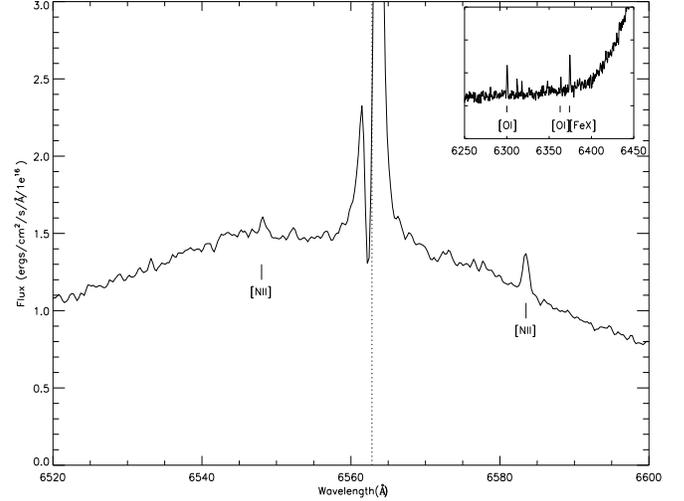, height=90mm, width=70mm, angle=90}
\caption[]{Intermediate resolution WHT/ISIS spectra of SN~2007rt at 153 days post-explosion, centered on the H$_{\alpha}$
region. Note the narrow P-Cygni profile not detected in the low
resolution spectra presented in Fig.~\ref{ha}. A number of narrow metal line
features were also detected in this wavelength range: [N {\sc ii}] 6548, 6583 \AA, [O
{\sc i}] 6300, 6343 \AA\ and [Fe X] 6374 \AA\  (see inset).}
\label{hahr} 
\end{figure}

To resolve this narrow H$_{\alpha}$ feature we obtained intermediate resolution spectra of SN~2007rt on the 4$^{\rm th}$ February 2008 with ISIS on the WHT. This revealed a previously unresolved P-Cygni profile (see Fig.~\ref{hahr}). The \fwhm~of the emission and absorption feature are 84 and 54 \kms, respectively (Table~\ref{narrow}). The blue wing of the absorption profile extends out to 128 \kms. If we assume that the narrow feature is representative of unshocked CSM, this edge velocity of 128 \kms corresponds to the terminal velocity of the wind of the progenitor star (see Sect.~\ref{progenitor} for more discussion on the progenitor). 

\subsection{H$_{\alpha}$ and He {\sc i} evolution}
\label{haevol}
From the H$_{\alpha}$ profiles in the first panel of Fig.~\ref{ha}, we can see that the line is highly asymmetric and consists of multiple components. Typically, the H$_{\alpha}$ profile of Type IIn supernovae can be decomposed into three parts: broad, intermediate and narrow. These are normally attributed to shocked (intermediate) and unshocked (narrow) circumstellar matter and emission from the underlying supernova ejecta (broad). In SN~2007rt, the red wing of the broad feature appears to be less extended than the blue wing, and both the broad and intermediate features show some asymmetry. The extent of the blue and red wings decrease rapidly with time, showing a reduction in the {\sc fwhm} of the broad component, and hence a change in the ejecta. There is also a notable blueshift in the 240 day spectrum, which becomes more pronounced by day 273, and which is accompanied by an increased asymmetry or decrease in the emission in the red wing. In the latest epochs, between 475 and 507 days, the blueshift is still present and there is a remarkable lack of redshifted emission. This change in the line profiles is accompanied by a simultaneous decline in the light curve of SN~2007rt (see Fig.~\ref{phot2}).  Asymmetry in the line profile and a blue-shifted peak are often related to dust formation. Since this affects the broad component of the line it could indicate the presence of newly formed dust in the SN ejecta \citep[viz. SN~1987A; see][]{dan89,lucy89}.

\begin{figure}
\epsfig{file=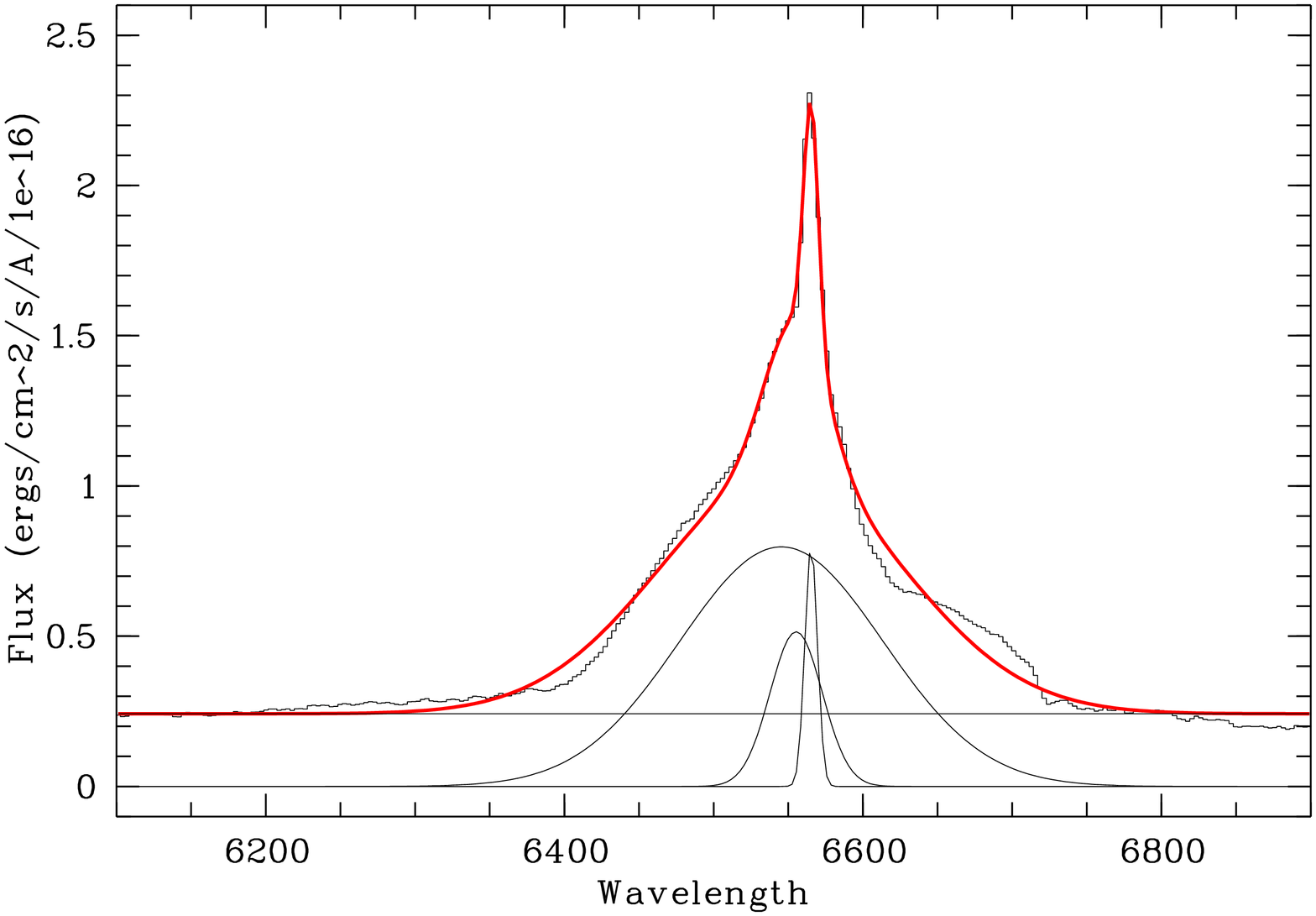, height=90mm, width=65mm , angle=-90}
\epsfig{file=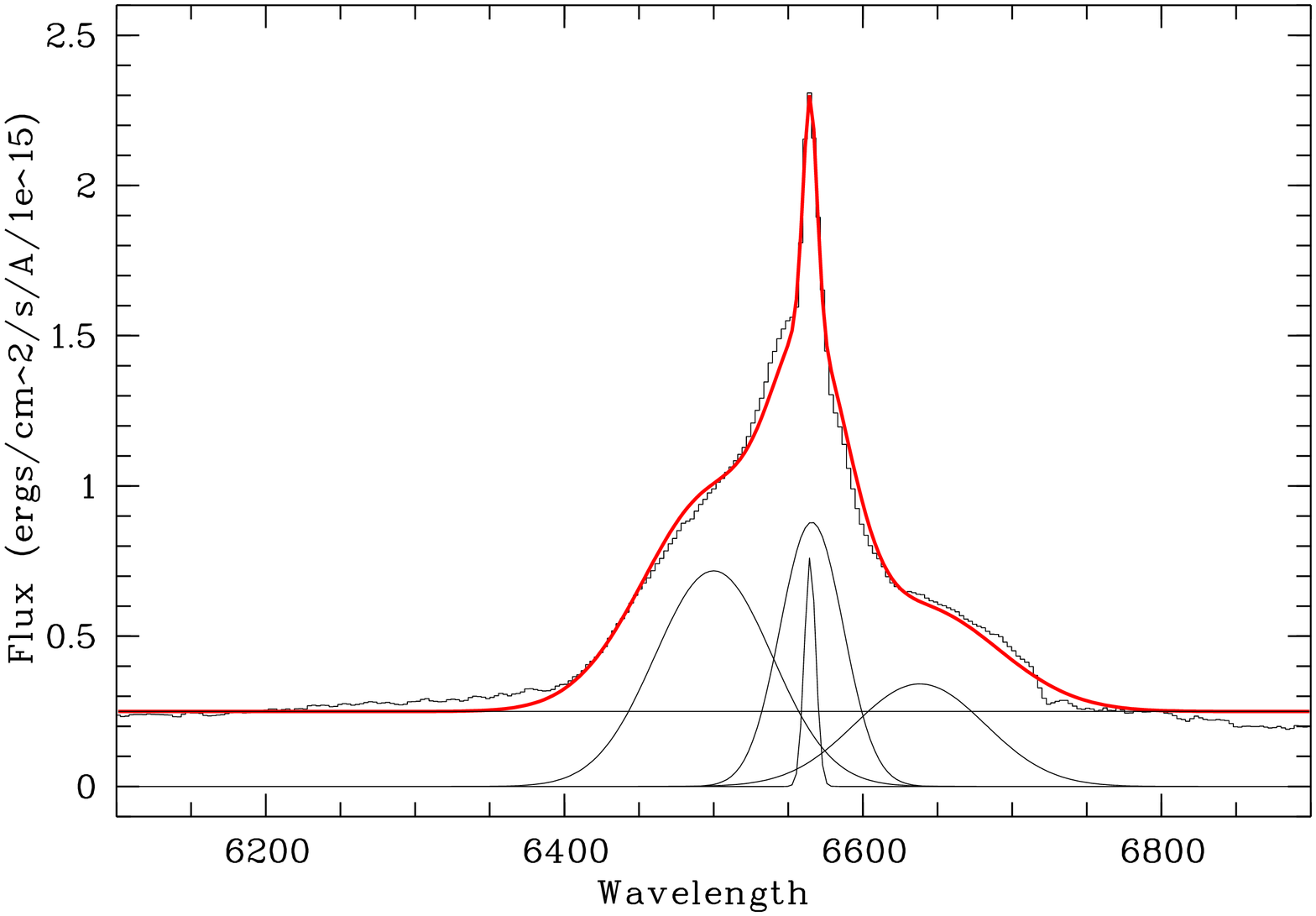, height=90mm, width=65mm , angle=-90}
\epsfig{file=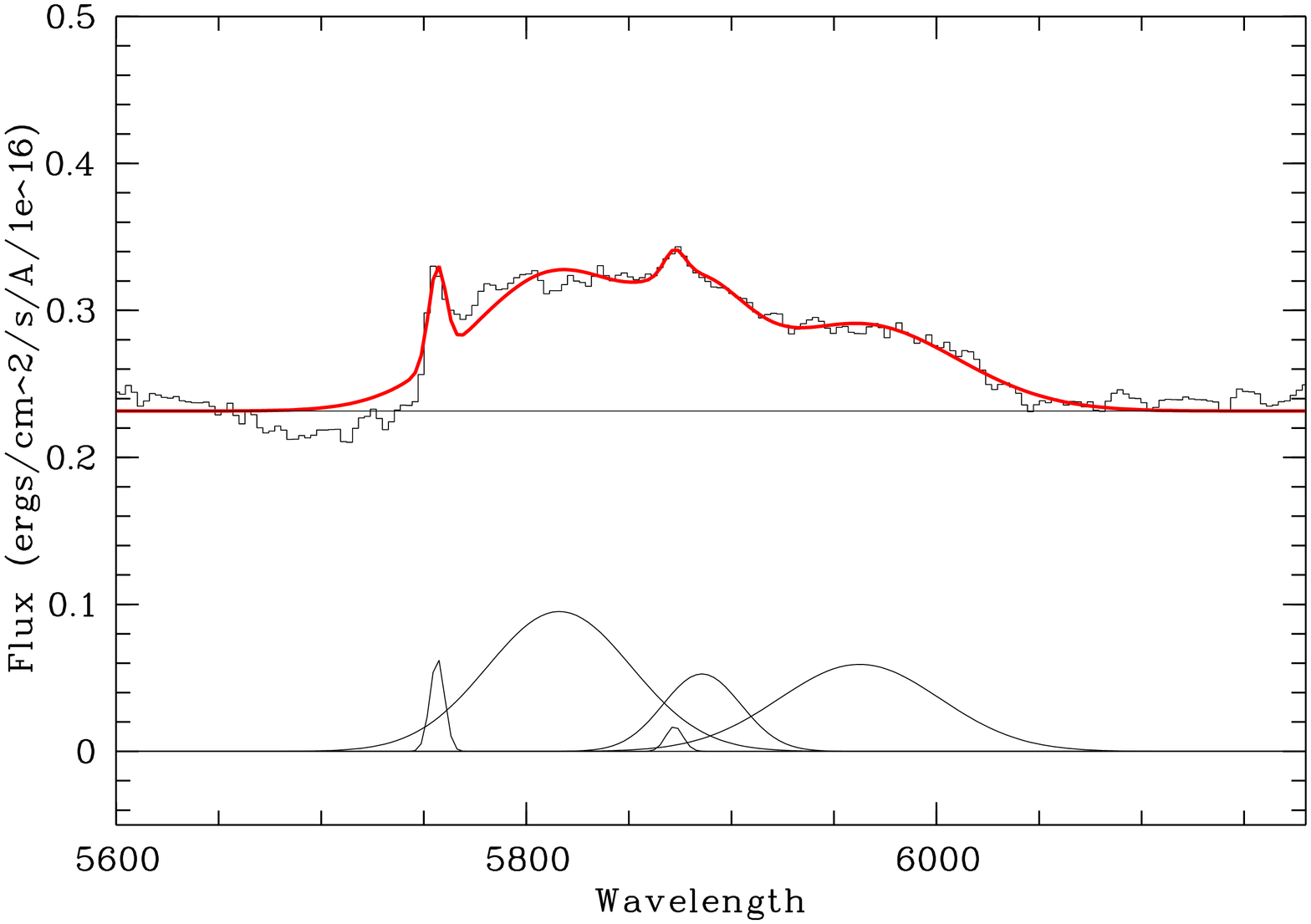, height=90mm, width=65mm , angle=-90}
\caption[]{Gaussian profile fits to H$_{\alpha}$ and He {\sc i} 5875 \AA\ line from the day 131 spectrum of SN~2007rt. {\bf Top Panel:} H$_{\alpha}$ line is reproduced using a narrow, intermediate and broad component. Note the poor fit to the wings, in particular in the red part of the line. {\bf Middle Panel:} Here 4 Gaussians are used to fit the H$_{\alpha}$ line; a narrow and intermediate component and 2 higher velocity (broad) components shifted bluewards and red-wards relative to the intermediate and narrow features. {\bf Lower Panel:} He {\sc i} 5875 \AA\  line from the day 131 spectrum fit with the same 4 Gaussians.  Note that in addition to the multiple components of He {\sc i} the left wing of the line is contaminated by narrow [N {\sc ii}] 5755 \AA\ line.}
\label{hefit} 
\end{figure}

\begin{figure}
\hspace{-5mm}\epsfig{file=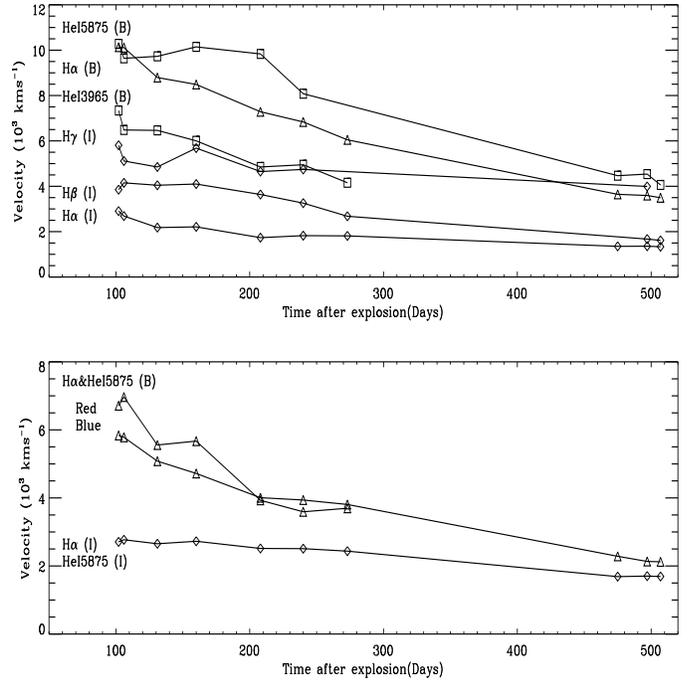, height=95mm, width=95mm, angle=90}
\caption[]{Line velocities as a function of time after assumed explosion. {\bf Upper panel}: velocities of He {\sc i} 3965, 5875 \AA, and H$_{\gamma,\beta}$ using one Gaussian for the broad component as presented in Table~\ref{hafittbl}. The broad and intermediate velocities of H$_{\alpha}$ are plotted as determined using 3 Gaussians.  The symbols represent: $\square$ - broad He, $\triangle$ - broad H, $\diamond$ - intermediate H.  {\bf Lower panel:} velocities of the intermediate ($\diamond$ ) and broad ($\triangle$) blue and red shifted components of the 4 Gaussian fits to the H$_{\alpha}$ and He {\sc i} 5875 \AA\ lines. The B \& I in the parenthesis indicate the broad and intermediate component velocities, respectively. Note that in this latter fit the He and H components have the same velocities. }
\label{vels} 
\end{figure}

In the first instance the complex profile of  H$_{\alpha}$  in SN~2007rt was fit with three Gaussian profiles allowing all parameters to be freely fit. In the early epochs, the He {\sc i} 5875 \AA\ line did not require an intermediate component and hence was fit only with components representing the broad and narrow features. An additional narrow component was added to account for the contribution to the profile from [N {\sc ii}] 5755 \AA.  In the case of He {\sc i} 5875 \AA\ the central wavelength of the narrow component was fixed.  The narrow line is no longer present in either the H or He lines after 208 days, and hence the profiles at these epochs were fit with only two Gaussians (see Fig.~\ref{hafitonline1}\footnote{Fig.~\ref{hafitonline1} is only available online}).  From day 475, two Gaussians were fit to the He {\sc i} 5875 \AA\ line representing the broad and intermediate components. The top panel of Fig.~\ref{hefit} shows a typical fit to H$_{\alpha}$ for one epoch. The parameters determined from this process for the intermediate and broad components are reported in Table~\ref{hafittbl}, those of the narrow components are presented separately. Uncertainties in the central wavelengths, FWHM, and  intensities of the fits are $\pm$ 5-7 \AA, 10\%, and 20\%, respectively. The intermediate component has a {\sc fwhm} velocity range of $\sim$2900 - 1800 \kms~over the first 192 days from discovery. There is a gradual reduction in the {\sc fwhm} of the broad components of the two lines over the observed period. In the first spectrum the broad component has a {\sc fwhm} velocity of $\sim$10,000 \kms, decreasing to 6000 \kms~171 days later  (see upper panel of Fig.~\ref{vels}). The fact that there is nearly a 40\% reduction in the width of the line clearly shows that this broad component is related to the ejecta. Nevertheless, the high velocity of 10,000 \kms~detected over three months after explosion is inconsistent with non-interacting CCSNe and requires an extremely energetic explosion. We note here that the high expansion velocities are only inconsistent if the age of SN~2007rt is greater than a month or two. However as discussed in Sect.~\ref{age} we cannot provide an accurate age estimate of the SN, the implications of these assumptions will be discussed further in Sect.~\ref{asymm}.
\begin{center}
\begin{table}
\caption[Decomposition fits to H$_\alpha$]
{\label{hafittbl2} Parameters of the multiple Gaussian fits to the H$_{\alpha}$ and He {\sc i} 5875 \AA\ lines, using four gaussian profiles (2 broad, 1 intermediate \&  1 narrow). \vspace{-4mm}}
\begin{flushleft}
\centering
\begin{tabular}{llccc} \hline \hline
Line                &   Day  &      $\lambda$  & FWHM     & Intensity \\
                       &            &      \AA               &  \kms      & 10$^{-15}$erg.s$^{-1}$cm$^{2}$\\ 
\hline 
He {\sc i} 5875  &102  & 5805.66  &  5838 &  11.31 \\  
Broad Blue        &106  & 5797.95  &  5782 &   8.95  \\ 
                            &131  & 5809.73  &  5081 &    8.86 \\ 
                            &160  & 5812.44  &  4720 &    8.23  \\
                            & 208 & 5811.79  &  4007 &    4.66  \\ 
		         & 240 & 5816.00  &  3939 &    4.58 \\ 
                            & 475 & 5822.00  &  2285 &  1.77 \\ 
                            & 497 & 5819.72  &  2134 &  0.83 \\
                            &507 & 5821.83 & 2123 & 0.82 \\

\hline
He {\sc i} 5875  & 102 & 5869.19 &  2708 &  2.65  \\  
Intermediate      &106  & 5864.20 &  2771 &  2.71  \\ 
                             & 131 & 5875.50 &  2653 &  3.12  \\ 
                             & 160 & 5876.78 &  2726 &  2.14  \\
                             & 208 & 5875.50 &  2515 &  2.96  \\ 
		          & 240 & 5875.97 &  2509&   2.95 \\ 
                             & 475& 5870.52&  1688 &  1.00 \\ 
                             & 497 & 5869.66 &  1703 &  1.00 \\
                             & 507 & 5871.27 &  1691& 0.99\\

\hline
He {\sc i} 5875  & 102  & 5943.84&  6712 &  10.65  \\  
Broad red           & 106  & 5936.13 &  6966 &   9.66  \\ 
                            & 131  & 5947.91  &  5556 &   6.61  \\ 
                            & 160  & 5950.42   &  5670&   7.88  \\
                            & 208  & 5949.97  &  3930 &  3.90  \\ 
		         & 240  & 5953.43  &  3593 &  2.85 \\ 
\hline
H$_{\alpha}$ 6563& 102  & 6500.00   &  5838 &  69.62 \\  
Broad blue              & 106  & 6500.00   &  5782 &  78.98 \\ 
                                  & 131  & 6500.00   &  5081 &  78.22\\ 
                                  &160  & 6500.00    &  4720 &  85.95  \\
                                  & 208 & 6500.00    &  4007 &  56.47 \\ 
                                  &240 & 6500.00     &  3939&  67.47 \\ 
                                  & 273 & 6500.00    &  3808 &  53.66 \\
                                  &475& 6500.00     &  2285 &  16.84 \\ 
                                  &497 & 6500.00    &  2134 &  9.72 \\
                                   &507 & 6500.00 & 2123 & 10.59\\
\hline
H$_{\alpha}$ 6563& 102 & 6563.53  &  2708 &  32.01 \\  
Intermediate            &106  & 6566.25  &  2771 &  39.45 \\ 
                                   & 131 & 6565.77  &  2653 &  51.14 \\ 
                                    & 160  & 6564.34&  2726 & 71.04  \\
                                   &207  & 6563.71 &  2515 & 85.36 \\ 
                                   & 240  & 6559.97 &  2509 &119.14  \\ 
                                   &273 & 6555.55  &  2436 &108.15  \\
                                   &475  & 6548.88 & 1688 & 64.16 \\
                                   &497 &6549.72 & 1703 & 41.66 \\
                                   &507 &6549.44& 1691 & 45.07\\
\hline
H$_{\alpha}$ 6563& 102 & 6638.18 &  6712 &  38.64  \\  
Broad red                & 106 & 6638.18 &  6966 &  47.82  \\ 
                                  &131  & 6638.18 &  5556 &  41.83  \\ 
                                  & 160  & 6638.18 &  5670&  55.25 \\
                                  & 207 & 6638.18 &  3930 &  35.68  \\ 
                                  & 240  & 6638.18 &  3593&  35.80  \\ 
                                  & 273  & 6638.18 &  3697&  22.10  \\ 
\hline
\multicolumn{5}{l}{Only parameters of intermediate and broad components}\\
\multicolumn{5}{l}{are presented here. Narrow components are in Table~\ref{narrow}.} \\
\end{tabular}
\end{flushleft}
\end{table}
\end{center}

Careful inspection of Fig.~\ref{hefit} reveals that the combination of the three Gaussian profiles does not provide a suitable fit to the red wing of the H$_{\alpha}$ profile. In addition, the peculiarly flat-topped He {\sc i} mentioned in Sect.\ref{specevol} is not consistent with a single broad component. An enhanced fit to both lines was found by fitting up to four Gaussian profiles: an intermediate component, broader blue and red-shifted components, and where appropriate, a narrow component (see Fig.~\ref{hefit}). The significance of these fits will be discussed in more detail in Sect.~\ref{discuss}, but briefly the two components at rest represent the shocked and unshocked circumstellar material. The other two components represent blobs of higher velocity material moving away from the supernova, one in the direction of the observer and one opposite to this.  To produce these fits, the H$_{\alpha}$ line was fit with multiple components, fixing the central wavelengths of the blue and red -shifted broad components in an iterative manner, whilst the other parameters were set free. For the helium line, the \fwhm~and separation of the broad and intermediate components determined for the H$_{\alpha}$ line, were used to fix their counterparts in the He {\sc i} feature. The lower panels of Fig.~\ref{hefit} show the resultant fits to H$_{\alpha}$ and He {\sc i} 5875 \AA. (Fits to the spectra for additional epochs can be seen in Fig.~\ref{hafitonline2}, which is only available online). The parameters for the intermediate and broad components are presented in Table~\ref{hafittbl2}. In the first epoch the intermediate component in both profiles has a \fwhm~of $\sim$2700 \kms, whilst the blue and red-shifted broader components are at $\sim$5800 and 6700 \kms, respectively (see Table~\ref{hafittbl2} and lower panel of Fig.~\ref{vels}). Over the first six months of the observational period of SN~2007rt, the broad component shows a 40\% reduction in \fwhm, whereas the intermediate component does not experience any noticeable change. It is therefore clear that the broader components are related to the ejecta, whilst the intermediate component is consistent with shocked material. In the last epochs taken a further 6 months later, only two components (a blue-shifted broad and intermediate component) were used to fit the data due to a lack of emission in the red wing of the lines.
\subsection{H$_{\beta}$}
\label{beta}
A strong intermediate component of H$_{\beta}$ is detected in SN~2007rt, with some indication of a weak narrow component in the earliest epochs (see second panel of Fig.~\ref{ha}).   Unlike H$_{\alpha}$, the higher order Balmer lines, H$_{\beta, \gamma}$, do not appear to have a broad component. For H$_{\beta}$ this is clearer, as it has a higher intensity relative to the strong blue continuum, but the behaviour of these lines are similar. The line profile of H$_{\beta}$ has an emission peak with a \fwhm~of $\sim$4000 \kms~in the first epoch of data. This corresponds to the intermediate component seen in H$_{\alpha}$, and changes at a similar rate (see Fig.~\ref{vels}).  In SN~1988Z, similar differences in the H$_{\alpha}$  and H$_{\beta}$ profiles were observed, with no broad component detected in the latter \citep{tur93}. Blended with the H$_{\beta}$ line is an absorption trough or depression in the continuum. Whilst there may be some contribution from hydrogen, the \fwhm~is twice the width of the emission in H$_{\beta}$, and it is unlikely to be due to hydrogen alone. 
\begin{center}
\begin{table}
\caption[Decomposition fits to H$_\alpha$]
{\label{narrow} Parameters of the Gaussian fits to the narrow lines detected in SN~2007rt  from the day 153 and 160 spectra.}
\begin{flushleft}
\centering
\begin{tabular}{llccc} \hline \hline
                    &        &      $\lambda$     & FWHM     & I \\
line             & Day&  \AA   &  \kms      &10$^{-15}$erg.s$^{-1}$cm$^{2}$\\ 
\hline 
\\
$[$Fe {\sc vii}] 6086 & 153  & 6087.26 &   43       &  0.079 \\  
$[$O {\sc i}] 6300      & 153  & 6300.63 &   47       &  0.090 \\  
$[$Fe {\sc x}] 6374   & 153  & 6374.80&   47        &  0.110 \\  
$[$N {\sc ii}] 6548     & 153  & 6548.35&   20        &  0.058 \\  
$[$N {\sc ii}] 6583     & 153  & 6583.72 &   29       &  0.157 \\  
H$_{\alpha}$ 6563 (e)&153 & 6562.43 &   84     &  8.84 \\
Narrow                 ~ (a) & 153  & 6563.89 &    54   &   3.41 \\
~~~~~~~~~~~~~~~(e)& 160  & 6561.80&  213 (463)&   3.97  \\
$[$Ne {\sc iii}] 3866  & 160  & 3865.03 &   289 (756)  &  0.596 \\  
$[$O {\sc iii}] 4363    & 160  & 4358.23 &   198 (650)  &  0.317 \\  
$[$O {\sc iii}] 4959    & 160  & 4956.10 &   ~~~~(568)  &  0.187 \\  
$[$O {\sc iii}] 5007    & 160  & 5008.36 &  ~~~~(537)     &  0.480 \\  
N {\sc ii} 5755          & 160& 5754.09  &    ~~~~(407)       &0.47 \\
He {\sc i} 5875          & 160  & 5873.27 &   ~~~~(396)     &   0.08 \\   
He {\sc i} 7065          & 160  & 7066.50 &   185 (424) &  0.252  \\                   
\hline
\multicolumn{5}{l}{The fwhm presented in parentheses are not corrected for}\\
\multicolumn{5}{l}{ resolution, all other values are.}\\
\multicolumn{5}{l}{e- emission, a- absorption}\\
\end{tabular}
\end{flushleft}
\end{table}
\end{center}

\subsection{Narrow emission lines}\label{narrowdis}

Besides the prototypical narrow H$_{\alpha}$ feature, a number of other narrow emission features, many of which are forbidden lines, were observed in the low resolution spectra of SN~2007rt. These narrow lines have been identified as He {\sc i} 5875, 7065 \AA, [O {\sc iii}] 4363, 4959, 5007 \AA, [N {\sc ii}] 5755 \AA, [Ne {\sc iii}] 3866 \AA, and very weak [Fe {\sc iii}] 5270 \AA\ and [Fe {\sc vii}] 6086 \AA. There is also a possible detection of [Fe {\sc xi}] 7891, although fringing in that part of the spectrum makes this identification marginal. Unfortunately the narrow lines were not resolved in our low resolution spectra. In the highest resolution spectrum in this low resolution group, taken on the 11$^{\rm th}$ February 2008 the \fwhm~of the lines are typically a few 100 \kms (see Table~\ref{narrow}). The [O {\sc iii}] lines have velocities of $\sim$200 \kms, which compares well with the narrow emission component of H$_{\alpha}$, suggesting they are produced in the same region of the CSM.  An intermediate resolution spectrum of the H$_{\alpha}$ region of SN~2007rt was obtained on the 4$^{\rm th}$ February 2008 with ISIS/WHT and revealed additional narrow features of [O {\sc i}] 6300, 6363 \AA, [N {\sc ii}] 6548, 6583 \AA\ and the high ionisation line [Fe {\sc x}] 6374 \AA. The [Fe {\sc vii}] 6086 line was also clearly detected in this spectrum. In Table~\ref{narrow} the parameters of the gaussian fits to the narrow lines detected in the TNG spectrum on 11$^{\rm th}$ February 2008 (day 160) and the ISIS/WHT spectrum (day 153) are presented as a guide to their parameters. Note that the velocities of the lines from the ISIS/WHT spectrum are significantly lower than derived from the lines in the spectrum taken seven days later, on the 11$^{\rm th}$ February. This is due to the differences in the spectral resolution, as the lines in the low resolution spectra are unresolved. We note here that as the lines are unresolved there may be some unidentified contribution to these narrow emission lines from underlying H {\sc ii} regions.
\section{Discussion}
\label{discuss}
\begin{figure*} 
\begin{center}
\epsfig{file=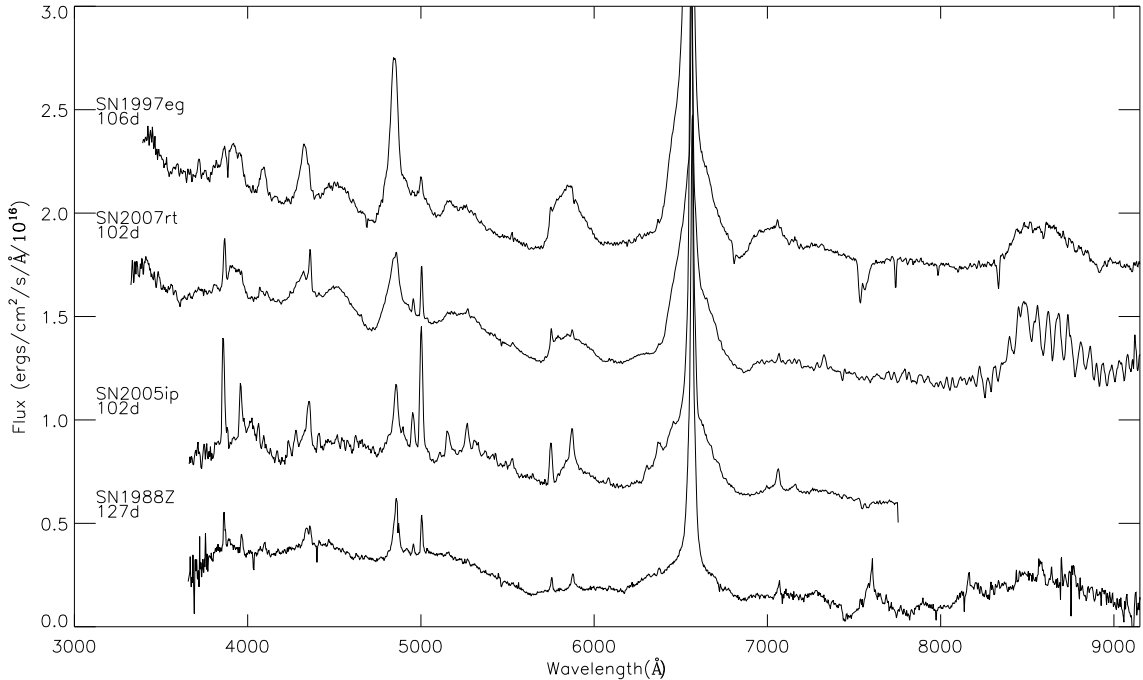, height=160mm ,angle=90} 
\caption[]{Comparison of SN~2007rt, SN~2005ip, the type IIn prototype, SN~1988Z ,and an additional Type IIn, SN~1997eg. Note the likeness of the underlying broad features of SN~2007rt, \& SN~2005ip, suggesting that these two supernovae are of the same age at the given epoch. The data for SN~2005ip and SN~1997eg are unpublished spectra from the Padova-Asiago Supernova Archive, and that of SN~1988Z is from \citet{tur93}.  }
\label{05ipspec}  
\end{center}
\end{figure*}
\begin{figure*} 
\begin{center}
\epsfig{file=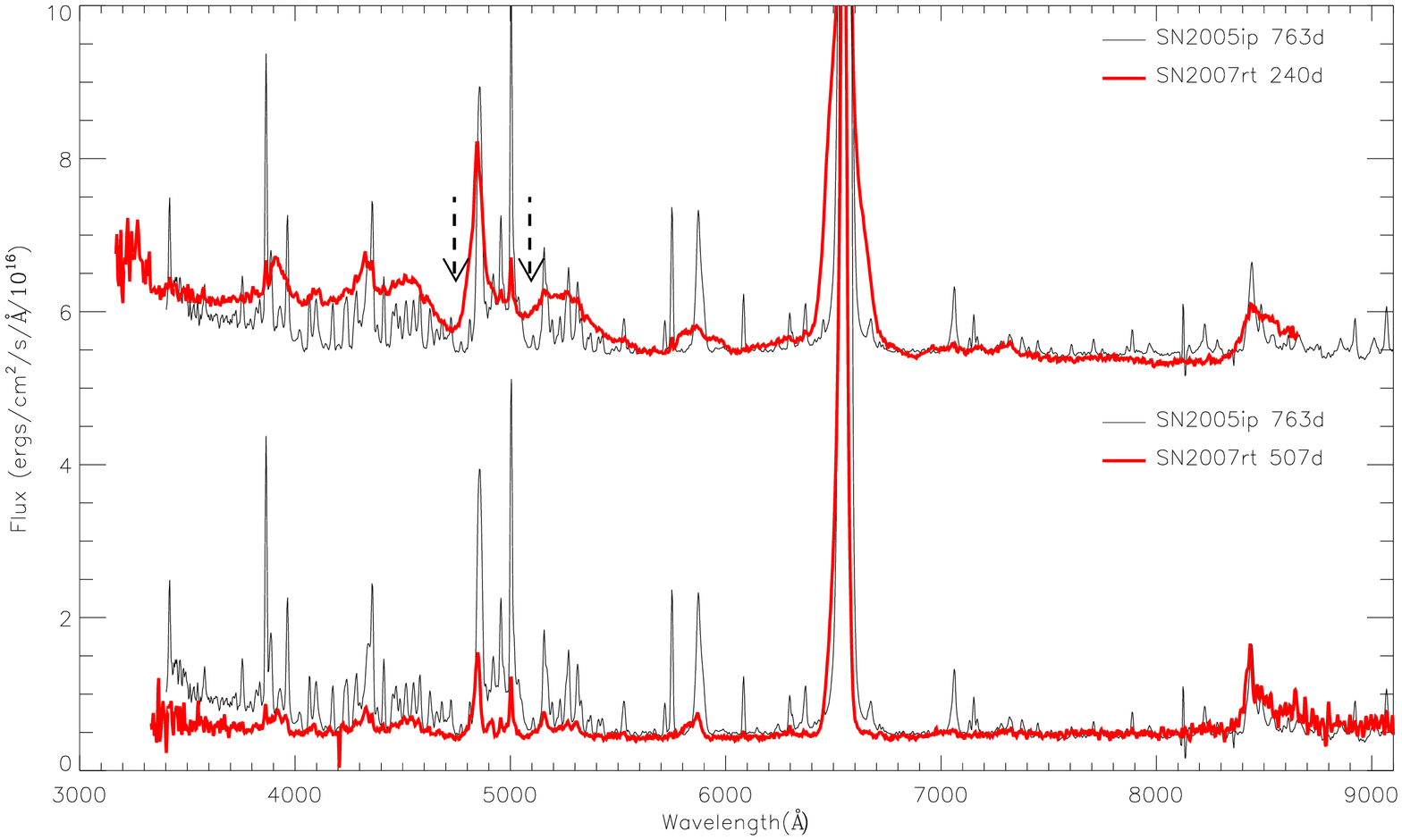, height=160mm ,angle=90} 
\caption[]{Comparison of SN~2007rt at 240 days and 507 post-explosion (thick red line) with SN~2005ip  at 763 days after explosion (thin black line). The data for SN~2005ip are unpublished spectra from the Padova-Asiago Supernova Archive.  }
\label{pseudo}
\end{center}  
\end{figure*}
In the introduction we mentioned the ambiguity surrounding the progenitors of Type IIn SNe and the debate over the mechanism for the SN~2002ic-like supernovae. The strong interaction in these so-called hybrid Type Ia/IIn objects makes it difficult to definitively conclude whether these are Type Ia or core-collapse supernovae. However in the case of SN~2007rt the presence of He in the ejecta, indicated by the broad He {\sc i} 5875 \AA\ line and the lack of broad forbidden Fe features at late phases, which are prominent in type Ia SNe, suggests that SN~2007rt is a core-collapse supernova.

\subsection{Comparison with other interacting SNe}
\label{ccsne}
In Sect.~\ref{photanalysis} we noted the similarity between SN~2007rt and another type IIn supernova, SN~2005ip. At early stages this SN displayed a weak series of both low and high ionisation, narrow features \citep[see their Fig. 2]{smith08b}, as detected here in SN~2007rt. SN~2005ip bears a strong resemblance to the prototype for the type IIn class, 1988Z, which also had a narrow, forbidden, emission line spectrum \citep{stat91, tur93}. In SN~2005ip and 1988Z there is a strong narrow He {\sc i} feature, with SN~2005ip developing an intermediate feature over time \citep[see their Fig. 8]{smith08b}.  SN~2007rt also has narrow He {\sc i} 5875 and 7065 \AA\ lines, which indicate the presence of helium in the circumstellar medium. However the main spectral difference between these two SNe is the presence of a broad He {\sc i} 5875 \AA\ feature in SN~2007rt.

This broad helium line is not usually detected in Type IIn spectra but is present in SN~2007rt from the first spectrum taken $\sim$102 days post-explosion and narrows with time.  A supernova with striking similarities to SN~2007rt is the type IIn SN~1997eg \citep{sal02,hoff08}. In Fig.~\ref{05ipspec} these two supernovae are compared along with SN~2005ip and SN~1988Z. SN~1997eg has strong broad He {\sc i} 5875 and 7065 \AA\ lines with velocities comparable to the broad H$_{\alpha}$ profile. \citeauthor{hoff08} suggest that the high He/H ratio in the ejecta of SN~1997eg, indicates that the progenitor star must have shed a significant amount of its hydrogen shell, possibly through an episode of mass-loss prior to the supernova explosion. This mass-loss episode would rid the star of a large quantity of its hydrogen shell and deposit the hydrogen rich material into its immediate environment, thus a helium rich atmosphere would be left behind in the progenitor star to be revealed in the supernova ejecta.  We suggest that this is also a likely scenario for SN~2007rt. However, it should be noted that strong lines can be formed not just by a high abundance but are dependent on the temperature and density conditions of the material.  Comparing the spectra of the four supernovae mentioned above, it is clear that there are varying degrees of He line strengths in these objects, with SN~2007rt being intermediate between SN~2005ip/SN~1988Z and SN~1997eg. One interpretation of this, is that there is varying degrees of H and He in the ejecta of these SNe. In addition to the high helium abundance, \citeauthor{hoff08} remarked that the peaked He profiles are suggestive of ejecta interacting with an asymmetric circumstellar medium, and this is substantiated by spectropolarimetry results.

At early times these type IIn supernovae have strong blue continua, with flatter red spectra. In the case of SN~2005ip a blackbody of 7300 K \citep{smith08b} was required to fit the data. Our earliest spectrum of SN~2007rt, at 102 days post-explosion, was poorly fit by a blackbody with $\sim$7550 K. Over time the blue excess decreased, and the red part of the spectrum flattened, making it difficult to fit a single blackbody. \citeauthor{smith08b} reported a similar effect in SN~2005ip. However, in this case the late time data resolved this continuum into a forest of high-ionisation (mainly Fe) emission lines. They suggest these lines are ionised by X-rays formed in the shocked material of a slow wind, and refer to a pseudo-continuum. This provides support for the suggestion that the blue excess in type IIn supernovae is related to the interaction of the ejecta with the CSM. \citeauthor{smith08b} suggest that the blue excess in SN~2006jc is a result of a similar effect, but the wind of the progenitor (likely a WR star) was faster, and hence the blue pseudo-continuum consisted of a blend of broader emission lines. If confirmed, this highlights the role the density and geometrical configuration of the progenitors wind plays in the evolution of these interacting SNe.

In the case of SN~2007rt the narrow H$_{\alpha}$ line suggests that the progenitors wind has a velocity similar to that of SN~2005ip ($\sim$100 \kms). The partially resolved features of O {\sc iii} at 200 \kms~suggest we should be able to at least partially resolve a forest of narrow emission lines, if present. Nevertheless even at 507 days post-explosion we do not see this forest of narrow emission lines in the blue, despite the broad features in the continua of the two SNe being similar.  The lack of evidence for this forest in our spectrum may be due to the continuum in SN~2007rt consisting of intermediate width emission lines, possibly of permitted/forbidden iron lines, formed in the shocked region. In Sect.~\ref{beta} we noted the depression in the continuum to the blue of the H$_{\beta}$ line. There is another such feature to the red of the [O {\sc iii} 5007] \AA\ line (see Fig.~\ref{pseudo}). These features are detected in many interacting supernovae such as 1997eg, 1995G, 2005la \citep[][see their Fig. 4]{sal02,hoff08,pas02,pas08b}, and coincide well with the broad Fe {\sc ii} absorption features from multiplet 42 (4924, 5018, 5169 \AA) present in type Ib/c spectra. However a comparison of late time spectra of SN~2007rt and SN~2005ip, as in Fig.~\ref{pseudo}, shows convincing evidence that in the spectra of interacting SNe this is due to a lack of emission lines amongst a dense forest, rather than an absorption trough.  This is also thought to be the case for SN~2006jc \citep{pas07}. The non-blackbody like continuum, similarity in the broad features in the blue part of the spectrum with those of many interacting supernovae, and the clear case of a pseudo-continuum in SN~2005ip leads us to believe that SN~2007rt also has a pseudo-continuum. However we can not rule out that a real thermal continuum contributes wholly or partially to the early time spectra of SN~2007rt.

\subsection{Properties of the CSM}
The CSM electron density and temperature can be inferred from the narrow forbidden lines in the spectrum of SN~2007rt. Coronal lines, such as [FeX] 6374 \AA,  as well as low ionisation [O {\sc iii}] lines are present, suggesting that there are regions of the CSM with different temperatures and densities. This is not unexpected in the complex environment of the CSM, and the detection of high ionisation Fe lines indicates that X-rays are produced in the shocked region. The presence of [Fe {\sc vii}] 6086 \AA\ and  [Fe {\sc x}] 6374 \AA\ combined with the absence of [Fe {\sc vi}] suggests that these lines are formed in a region of the CSM with an electron temperature between  2.5 and 8 $\times$ 10$^{5}$K \citep{bry08}.  The [O {\sc iii}] lines, however, are expected to form in cooler regions with electron temperatures of $\sim$1 $\times$ 10$^{5}$K \citep{bry08}. Using the relationship between the [O {\sc iii}] 
lines and electron density by \citet{ost89}, as cited in \citet{sal02}, we can determine the density of 
the CSM. Thus using the temperature range implied above, with the [O {\sc iii}] line intensity 
ratio (I$_{4959+5007}$/I$_{4363}$) of 2.93 the density of the CSM, where the [O {\sc iii}] lines form, is 4.6 $\times$ 10$^{7}$ cm$^{-3}$. 

The narrow H$_{\alpha}$ component is present in the early spectra of SN~2007rt but has completely disappeared in the spectrum taken on 1$^{\rm st}$ May 2008. If we assume that this is due to the ejecta having swept up the entire CSM, we can determine the extent of the CSM. Assuming the explosion date discussed in Sect.~\ref{photanalysis} and an ejecta velocity of 10, 000 \kms,  the unshocked region extends out to a maximum of 2.13 $\times$ 10$^{16}$ cm. Material ejected from a star at a wind speed of $\sim$100 \kms~would take $\sim$ 70 years to reach such a distance, hence the mass-loss must have occurred on this timescale. However, given the high densities of the CSM, as implied above from the O {\sc iii} lines, it is possible that this narrow H$_{\alpha}$ feature disappears due to recombination. This explanation has been given for the disappearance of such features in SN~1994aj and SN~1996L \citep{b98,b99}.

\subsection{Nature of the Progenitor Star}
\label{progenitor}
In Sect.~\ref{narrowha} we discussed the very narrow P-Cygni feature detected in SN~2007rt, the absorption component of which has a \fwhm~of 54 \kms, and extends out to 128 \kms. Narrow P-Cygni profiles have been detected in a number of type IIn SNe. However, they appear to fall into two different categories; those with very narrow profiles with the blue edge of the absorption profile extending out to $<$200 \kms~and those with velocities in the range 600-1000 \kms. Some supernovae which fall into the latter category are SNe~1994aj, 1994W, 1995G,  and 1996L, while those with lower velocities are SNe~1997ab, 1997eg, 2005gj \citep{chug94,sol98,b98,b99,pas02,sal98,sal02,tru08}. As mentioned in  Sect.~\ref{narrowha} these P-Cygni profiles provide insight into the stellar wind velocity of the progenitor star. The high velocities observed in SN~1994W and others are consistent with the wind velocities observed in WR stars \citep{crow07}. Luminous blue variables (LBVs) have slower wind velocities in the range of 100-500 \kms~\citep{stahl01,stahl03,smith07}, which may explain some of the objects in the low velocity category, such as in the case of SN~1997ab. Typically red supergiants have velocities of approximately 10 \kms~but there are a few cases for which 30-40 \kms~edge velocities have been detected \citep[viz. VY CMa see][and references therein]{smith09b}. Therefore it cannot be ruled out that the wind velocities of some of these lower velocity objects may be consistent with extreme red supergiants. The expansion velocity of $\sim$ 128 \kms~is on the lower extreme of the LBV range, comparable to quiescent LBVs such as HD160529 \citep{stahl03}, and is quite high for red supergiants.

Type II spectra show no indication of He spectral lines at optical wavelengths, except for at very early stages ($\sim$ 1 week) when the high temperatures in the ejecta allow for their formation. Type IIb are the only SNe which have He and H present in their spectra, with possibly type Ib showing a trace of H accompanied by He.  If the interpretation that the broad He {\sc i} 5875 \AA\ line in SN~2007rt is formed by a high helium abundance is valid, it suggests that the atmosphere of SN~2007rt's progenitor has a higher He/H ratio than that of many Type II SNe progenitors. This places the progenitor of SN~2007rt as a transitional object between those of normal Type IIP's and those of hydrogen stripped core collapse SNe. The combination of H and He in the ejecta would suggest either the progenitor passed through an LBV phase and has lost a significant amount of its H shell through its previous mass-loss history or that the progenitor is in a more evolved WN or WNH phase, i.e. mass-loss via stellar winds has revealed H-burning products at the stars surface but not He-burning products as would be the case in WC stars.   However, the very low (LBV-like) wind velocity  of the unshocked material is not consistent with a WR wind, as velocities in such stars are typically greater than 500 \kms~\citep[][and references therein]{crow07}. If the progenitor is a WR star, the CSM detected by the narrow H components could be the result of an LBV outburst, which occurred prior to the progenitor entering a WN phase.
In this case the progenitor would need to be in a very early stage of the WN phase, as otherwise the LBV wind would be swept up by the WN wind \citep{vm07}.  However we should note that there is no clear distinction between quiescent LBVs and WNH stars, as many of the latter are know to be quiescent LBVs \citep[viz. AG Car, see][and references therein]{smith08c}. 

\subsection{Explanation of the H and He {\sc i} Asymmetries}
\label{asymm}
As discussed in Sect~\ref{haevol}, the H$_{\alpha}$ and He {\sc i} 5875 \AA\  profiles in the spectra of SN~2007rt are  peculiar.  An asymmetry is present in these lines from the first spectrum onwards (see Fig.~\ref{ha}). In the last few epochs of our spectral dataset, from 240 days, a blueshift is detected. 

\subsubsection{Late phase evolution: $\sim$240-507 days post-explosion.}

Once the temperature in the ejecta has dropped below the threshold for dust grains to condense, dust can form. The presence of dust grains causes a net blueshift due to the absorption of redshifted light. Hence, asymmetric and blueshifted profiles formed as the ejecta expands and cools, can be explained by dust forming in the ejecta. In Fig.~\ref{ha} it can be seen that the asymmetry of the H$_{\alpha}$ profile in SN~2007rt increases with time and at late phases become blueshifted with significant absorption of the redshifted light. This behaviour appears to be consistent with the presence of newly formed dust. In addition there is an increasing asymmetry in the He {\sc i} line.  Evidence of such behavior due to dust has been seen in a number of Type II SNe, viz. 1987A \citep{dan89,lucy89}.  In SN~2007rt this blueshift is first detected in the spectrum taken on day 240 and becomes more pronounced by day 273. At even later epochs (475-507 days) the H$_{\alpha}$ and He {\sc i} 5875 \AA\  lines show significant absorption of the redshifted light. Additionally, there is a significant decrease in SN~2007rt's magnitude during this late phase.  The R-band magnitude declines at a rate of 0.01 mag d$^{-1}$ from 458 to 562 days post-maximum. In the case of Type II SN~1987A a clear rise in the IR magnitudes was detected and is accompanied by a decline in the optical band at over 450 days post-explosion, indicating  the formation of dust \citep[0.016 mag d$^{-1}$ at 467-562 days,][]{whit89}. Hence the formation of dust is the most probable justification for the decline in SN~2007rt's lightcurve in these latter points.

\subsubsection{Early phase evolution: $\sim$102-240 days post-explosion.}

The explanation of asymmetries in the H and He profiles in the earliest SN~2007rt spectra is uncertain, due in part to the uncertainty in its age. Here we outline two possible scenarios: (1) the object is young and can form dust in the fast expanding ejecta, (2) the broad components are inconsistent with the SNe's  age and it has an asymmetric or bipolar outflow. 

As mentioned above the reduced emission in the red wing of line profiles are suggestive of dust formation, however most dust detections have been made in late-time ($\sim$ 300 days) SNe spectra. The asymmetry in the first spectrum, taken approximately 102 days post-explosion, would appear to be inconsistent with the late-time formation of dust. Additionally if dust is invoked to explain the asymmetry in the profiles, an explanation for high ejecta velocities of 10,000 \kms at approximately 102 days after explosion is required due to the high energetic explosions implied by such velocities. Typically non-interacting core collapse supernovae have velocities of 10,000 \kms~and greater only within 30 days of the explosion \citep[see Fig. 5]{pat01}. At 100 days post-explosion, 7000-5000 \kms~are more typical expansion velocities. Whilst the age of SN~2007rt is uncertain it is unlikely to be less than one month old, as the SN was first observed 21 days after discovery, was caught post-maximum, and in an interacting SNe such as this the spectrum is expected to be significantly bluer than detected (see Sect.~\ref{age}). 

Nevertheless assuming that SN~2007rt has an age significantly less than or equal to 100 days, it is difficult to form dust at an early epoch. Unlike the case of the peculiar Ibn SN~2006jc \citep{sak07,mat08, smith08a,dicarlo08,noz08}, the dust must have formed in the ejecta not in a post-shock region as the asymmetries observed are in the broad rather than the intermediate component. The only other object, for which dust has been invoked to explain early-time fading of the broad component, is SN~2005ip \citep{smith08b}. \citeauthor{smith08b} suggest that the dust forms in the fast expanding ejecta, however it is unclear what mechanism would allow for dust formation in the high temperature gas of the ejecta. A possible added support to dust formation at a young age is the detection of IR excess by \citet{fox09} from NIR photometry of SN~2005ip from 50-200 days post-discovery. 

An alternative scenario requires the presence of an aspherical or bipolar outflow. This scenario does not require such high expansion velocities as is the case above and hence is more consistent with our adopted explosion epoch. It also provides fits to the H$_{\alpha}$ and He {\sc i}  5875 \AA\  lines, which are more consistent with their profiles. The profile fit in this case requires a blue and red-shifted component at velocities in the range 6000-7000 \kms~with an additional intermediate and narrow component (see Fig.~\ref{hefit}). The blue and red-shifted components represent high velocity material moving away from the SN, which may be indicative of a asymmetric outflow of material from the SNe. The profiles seen in SN~2007rt are reminiscent of the double-peaked profiles seen in Type Ib/c, viz. the broad-lined Ic, SN~2003jd, and the Type IIb, SN 2006T \citep{mae08, val08a,tau09}.  In the case of these latter objects the profile shapes can be explained by aspherical jet-like explosions viewed nearly sideways on \citep{maz05,mae08}.  \citeauthor{maz05} and \citeauthor{mae08} suggest that these double-peaked features can be detected if viewed from angles of 60-90 degrees to the jet axis. For SN~2007rt, there is still a significant amount of H in the shell and any model would need to account for this.

\section{Conclusions}

We have presented a photometric and spectroscopic analysis of the Type IIn supernova SN~2007rt over more than 1 year after discovery.  At 102 days post-explosion, SN~2007rt bears a striking resemblance to the type IIn supernovae SN~1988Z, SN~1997eg and SN~2005ip \citep{stat91,tur93,art99,sal02,hoff08,smith08b}, with strong narrow/intermediate/broad H${\alpha}$ emission lines, a strong blue continuum, as well as weak narrow emission lines from neutral to highly ionised states (viz. He {\sc i}, [O {\sc iii}], N {\sc ii}, [Fe {\sc vii}], [Fe {\sc x}]).  

The narrow H lines indicate the presence of a H-rich CSM surrounding the SN. An intermediate resolution spectrum of the H$_{\alpha}$ region resolved the narrow emission feature into a P-Cygni profile with an edge velocity of 128 \kms. This suggests the SN progenitor underwent mass-loss with velocities at the low end of those detected in LBV winds. The first spectrum contains a strong intermediate H$_{\alpha}$ component suggesting the ejecta had already begun to interact with the CSM. This is supported by the light curve of SN~2007rt, as it evolves very slowly over the early epochs of our data, declining at a rate of 0.003 mag.d$^{-1}$. By day 240 the narrow H$_{\alpha}$ component has disappeared and provides an estimate of the maximum extension of 2.13 $\times$ 10$^{16}$ cm for the unshocked CSM shell. Furthermore, a blue shift in the broad H$_{\alpha}$ feature at this late stage suggests that dust has begun to form in the ejecta.

A broad He {\sc i} 5875 \AA\ component was also present in SN~2007rt. This is not a typical feature of SNe type IIn, and may be indicative of ejecta with a higher He/H ratio than generally observed in type IIn SNe. It is therefore possible that its progenitor is transitional between those of normal type II's and hydrogen stripped core-collapse SNe, which as a result of its previous mass-loss history has lost a large amount of its hydrogen shell. There is also a hint of He present in the CSM, and hence the progenitor may be a WNH star in an early stage of evolution or an LBV which has lost a significant amount of H in  previous mass-loss events. Throughout the spectral observations, the H$_{\alpha}$ profiles show a strong asymmetry that increases over time, the red wing being dampened compared to the blue wing. The presence of dust in the ejecta beyond 240 days is clear, however what causes the asymmetry in the earlier spectra is less certain. Two possible scenarios are presented to account for this which cannot be distinguished by our current dataset: (1) the supernova is significantly  younger than estimated and dust is formed through some unknown mechanism in the fast expanding ejecta or (2)  an asymmetric or bipolar outflow viewed nearly side on accounts for the asymmetry in the early epochs. The first scenario is similar to that invoked for SN~2005ip by \citet{smith08b}, however the mechanism for forming dust in the fast expanding ejecta is unclear. The second scenario relies on the SNe being older than one or two months and that the expansion velocity behaves like normal non-interacting core-collapse supernovae.

\vspace{-0.3cm} 
\section{Acknowledgements}  
The authors are grateful for the feedback from the referee and  to Daniel Mendicini and Martin Nicholson for providing a number of photometric data points (http://ar.geocities.com/daniel\_mendicini/index.html; http://www.martin-nicholson.info/1/1a.htm). In addition we are grateful for the support from Avet Harutyunyan at the TNG, La Palma. CT acknowledges financial support from the STFC. FPK is grateful to AWE Aldermaston for the award of a William Penney Fellowship. This paper is based on observations from a number of telescopes; 2.2-m Calar telescope at the German-Spanish Astronomical Center at Calar Alto operated jointly by the Max-Planck-Institut f\"ur Astronomie (MPIA) and the Instituto de Astrof\'isica de Andaluc\'ia (CSIC), 1.82m Copernico telescope at Asiago Observatory operated by Padova Observatory, ALFOSC owned by the Instituto de Astrof\'isica de Andalucia (IAA) and operated at the Nordic Optical Telescope under agreement between IAA and the NBIfAFG of the Astronomical Observatory of Copenhagen, WHT and the Italian Telescopio Nazionale Galileo (TNG) operated by the Isaac Newton Group, and  the FundaciÑn Galileo Galilei of the INAF (Istituto Nazionale di Astrofisica) on the island of La Palma at the Spanish Observatorio del Roque de los Muchachos of the Instituto de Astrof\'isica de Canarias. Observations were also carried on with the 0.35-m SLOOH telescope at the Teide Observatory (Canary Islands, Spain). SLOOH (http://www.slooh.com) is a subscription-based website enabling 
affordable, user-friendly control of observatories in Teide, Chile, and Australia.

\vspace{-0.4cm}
\bibliography{sn07rt}  
\newpage
\appendix
\section{Sequence Stars in local region of SN~20007rt}
\begin{figure}[]
\epsfig{file=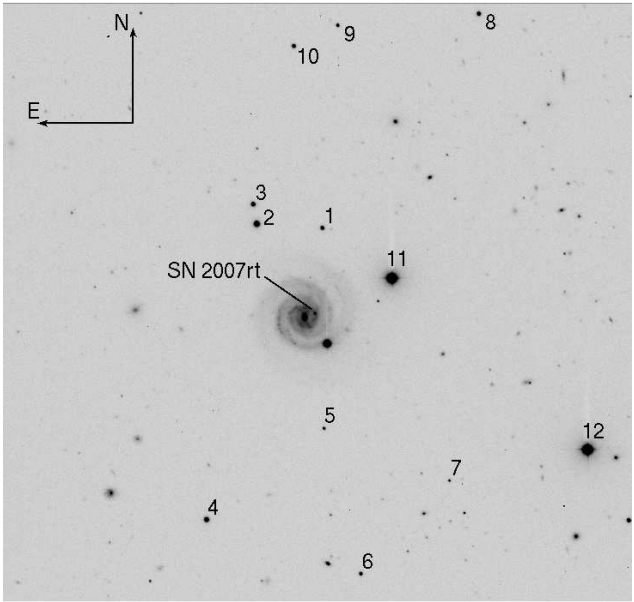,height=80mm}
\caption[]{R band image of the region of SN~2007rt. Taken at CAFOS at Calar Alto on 19$^{th}$ March 2009.}
\label{seq}
\end{figure}

The sequence stars used to estimate the SN magnitudes are identified in Fig~\ref{seq} and
their magnitudes are presented in Table~\ref{photseq}
\begin{flushleft}
\begin{table*}[]
\hspace{-30cm}\caption[Sequence stars magnitudes]
{\label{photseq} 
UBVRI magnitudes of the sequence stars in the field of SN~2007rt.}
\begin{flushleft}
\centering
\begin{tabular}{lcccccccccc} \hline \hline
 ID &    U & U$_{err}$ & B & B$_{err}$ & V & V$_{err}$ & R & R$_{err}$ & I & I$_{err}$ \\
\hline 
\\
1   & 18.037 & 0.003 & 18.225 & 0.002 & 17.669 & 0.004 & 17.294 & 0.000 & 16.905 & 0.008 \\ 
2   & 19.052 & 0.003 & 18.022 & 0.004 & 16.462 & 0.007 & 15.330 & 0.008 & 13.796 & 0.024 \\
3   & 20.243 & 0.022 & 19.269 & 0.016 & 17.785 & 0.005 & 16.652 & 0.009 & 15.215 & 0.010 \\
4   & 17.956 & 0.012 & 17.534 & 0.016 & 16.682 & 0.007 & 16.217 & 0.013 & 15.778 & 0.005 \\
5   & 20.730 & 0.014 & 20.083 & 0.017 & 19.177 & 0.007 & 18.673 & 0.003 & 18.200 & 0.003 \\
6   & 20.968 & 0.026 & 20.048 & 0.011 & 18.607 & 0.003 & 17.781 & 0.008 & 16.951 & 0.010 \\ 
7   & 17.541  & 0.026& 21.797 & 0.009 & 20.321 & 0.016 & 19.288 & 0.012 & 17.896 & 0.008 \\
8   & 17.623  & 0.039& 17.875 & 0.006 & 17.360 & 0.011 & 17.028 & 0.007 & 16.659 & 0.005 \\
9   &  21.031 & 0.006& 19.922 & 0.004 & 18.775 & 0.017 & 18.055 & 0.015 & 17.428 & 0.032 \\
10 & 17.756 & 0.002 & 18.012 & 0.007 & 17.549 & 0.009 & 17.247 & 0.006 & 16.974 & 0.029 \\
11 &               &            & 13.639 & 0.006 & 13.140 & 0.004 &               &             &              &            \\
12 &               &            & 13.792 & 0.007 & 13.064 & 0.009 &               &             &              &            \\
\hline
\\
\end{tabular}
\end{flushleft}
\end{table*}
\end{flushleft}

\Online
\begin{figure*}
\epsfig{file=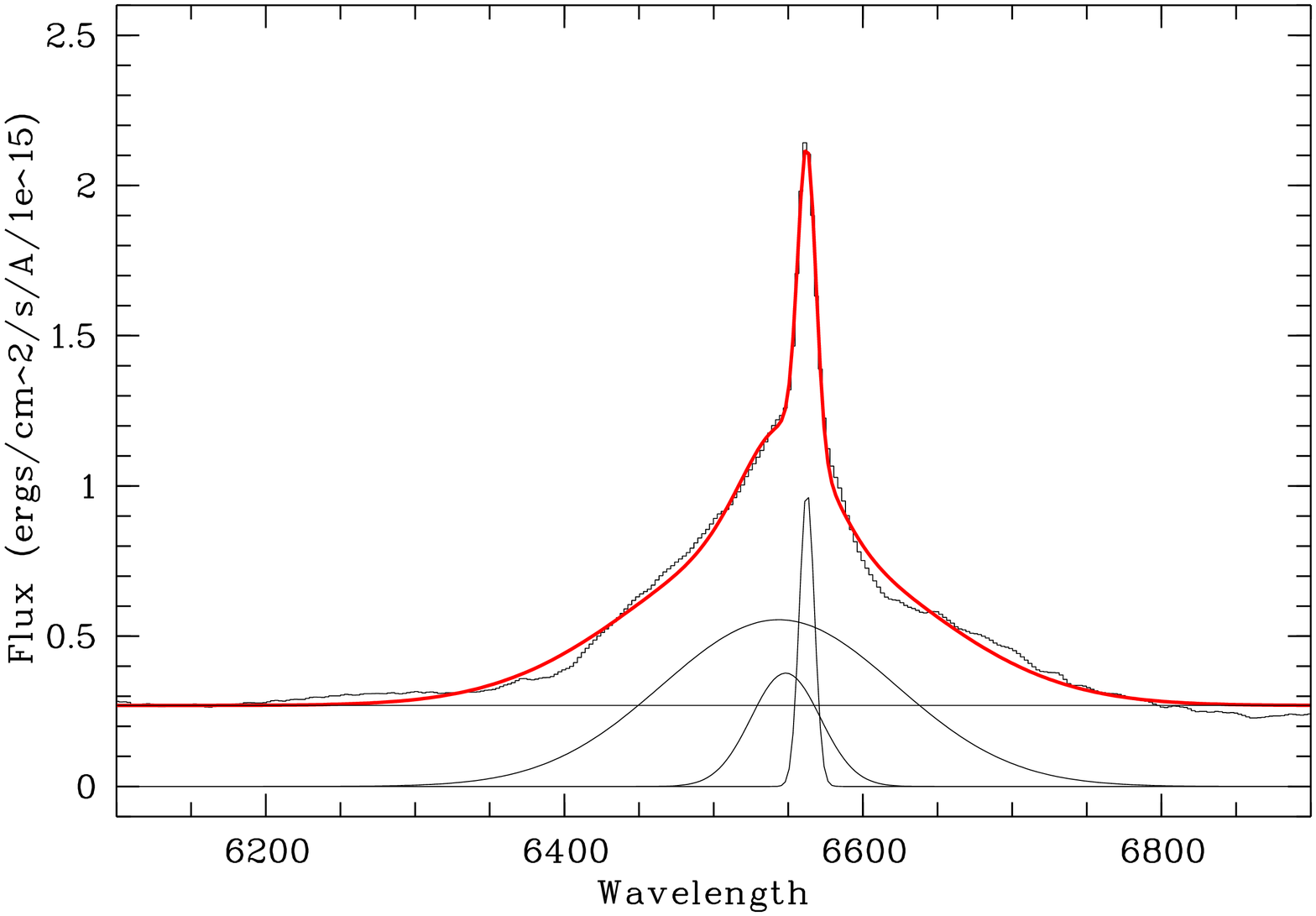, height=90mm, width=62mm , angle=-90}
\epsfig{file=./figures/ha_20080113.ps, height=90mm, width=62mm , angle=-90}
\epsfig{file=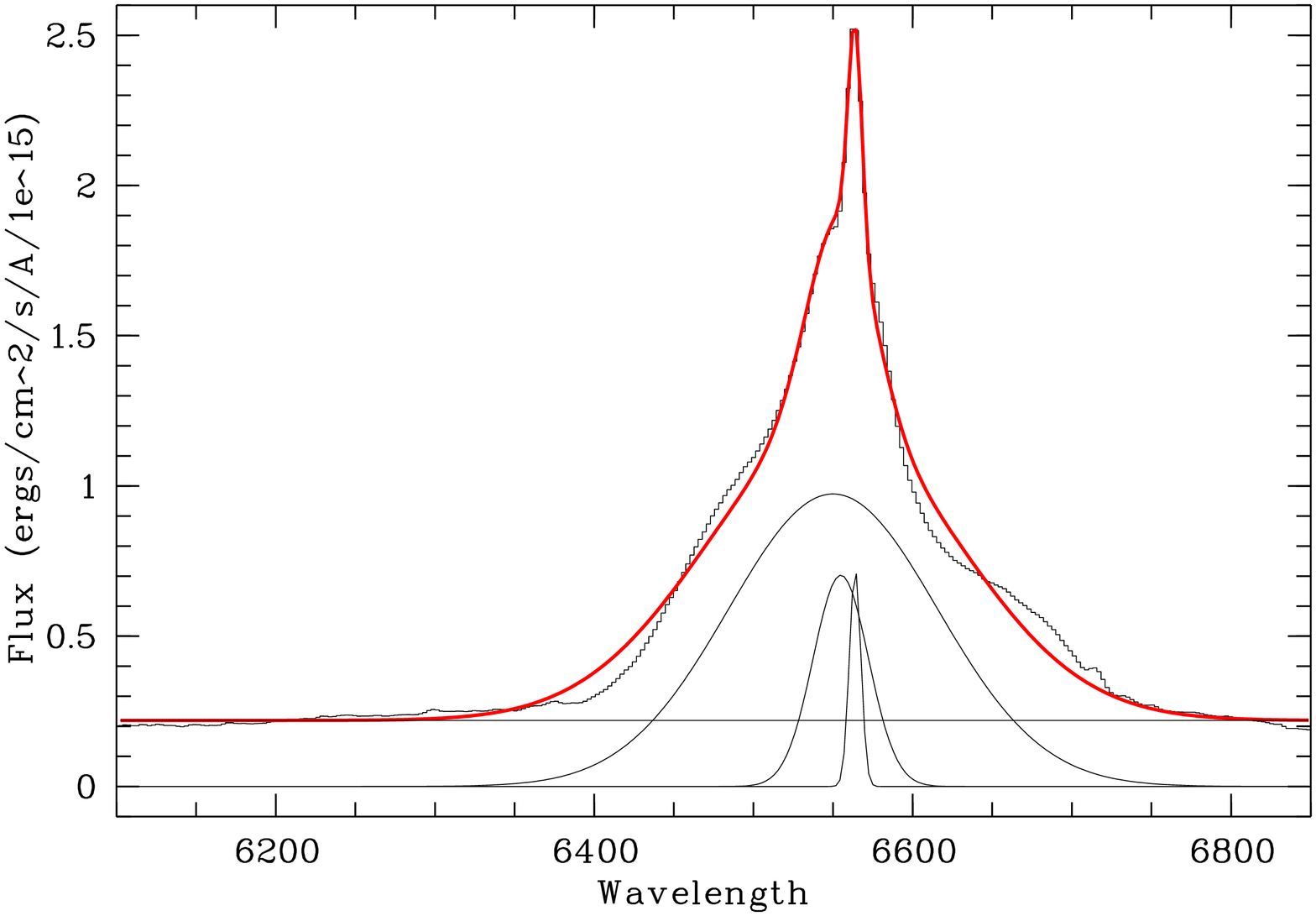, height=90mm, width=62mm , angle=-90}
\epsfig{file=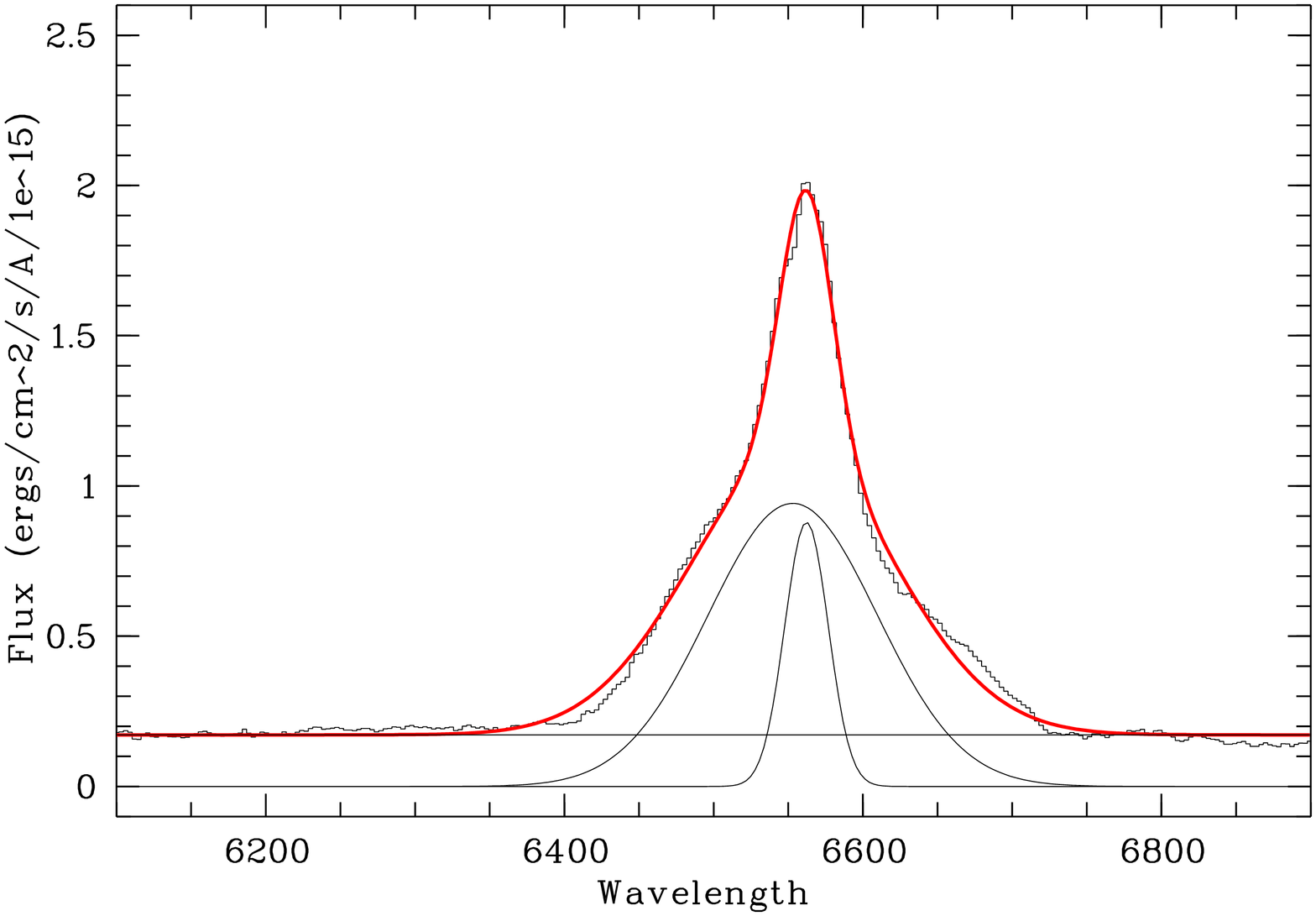, height=90mm, width=62mm , angle=-90}
\epsfig{file=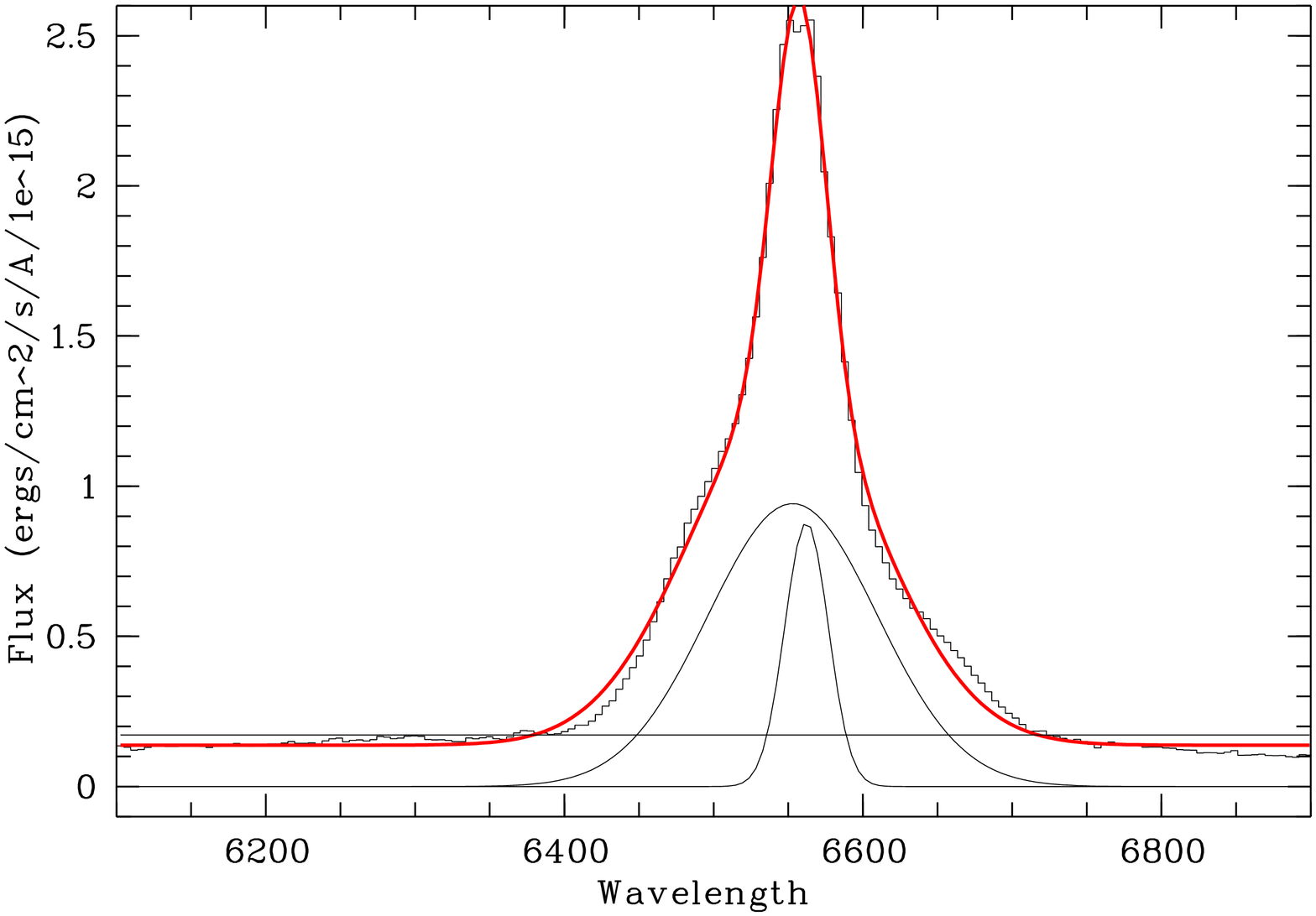, height=90mm, width=62mm , angle=-90}
\epsfig{file=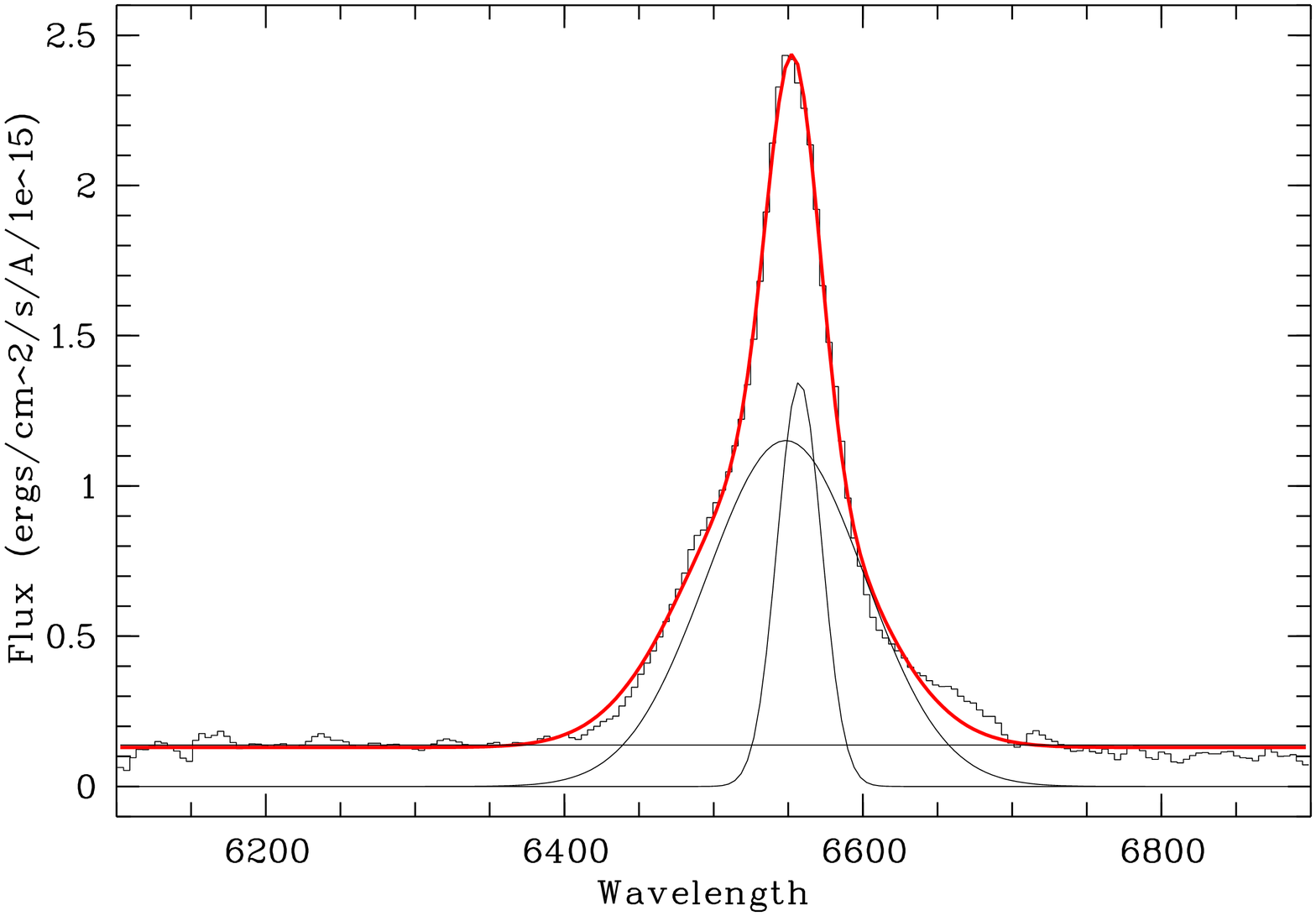, height=90mm, width=62mm , angle=-90}
\epsfig{file=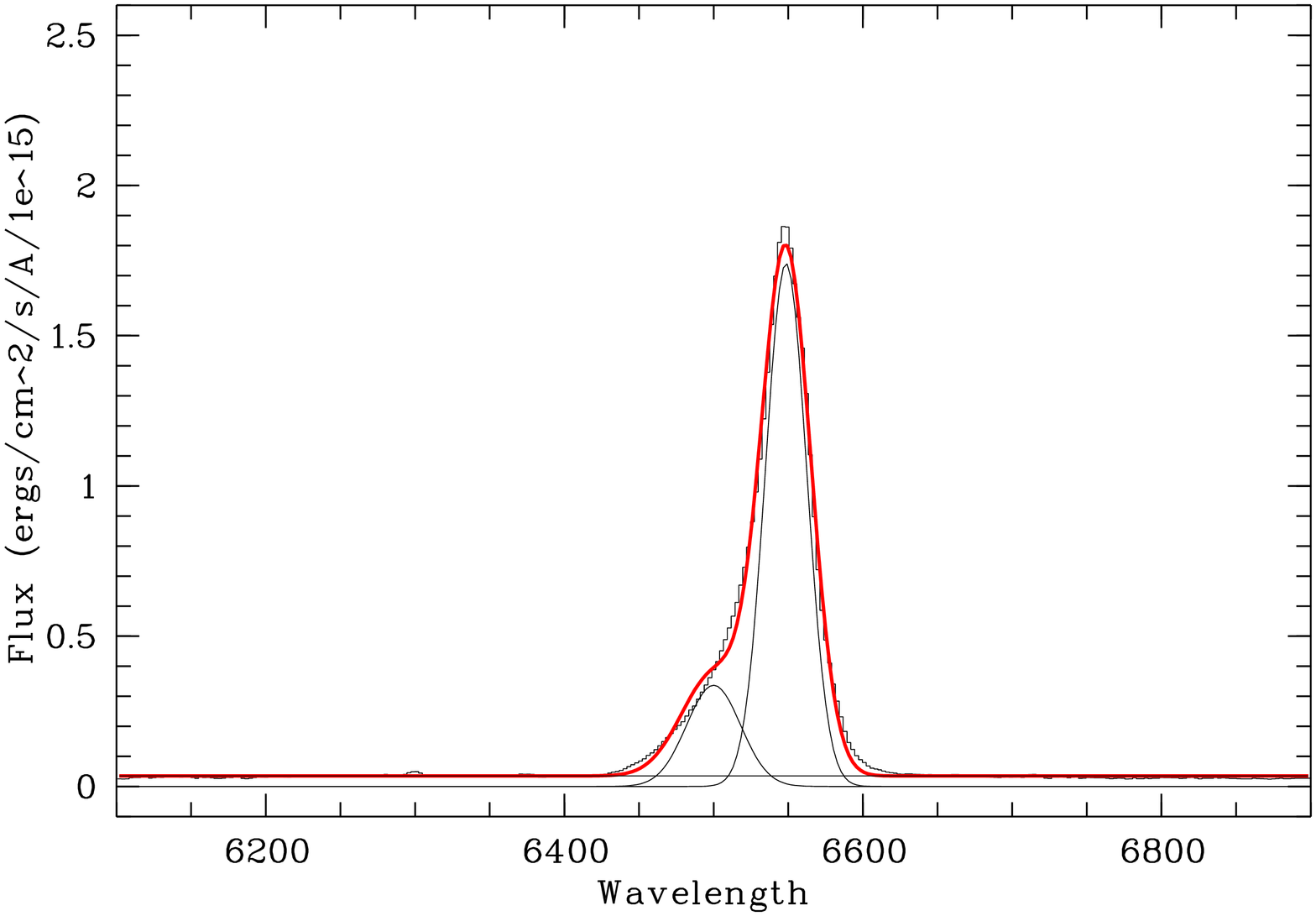, height=90mm, width=60mm , angle=-90}
\hspace{3mm}\epsfig{file=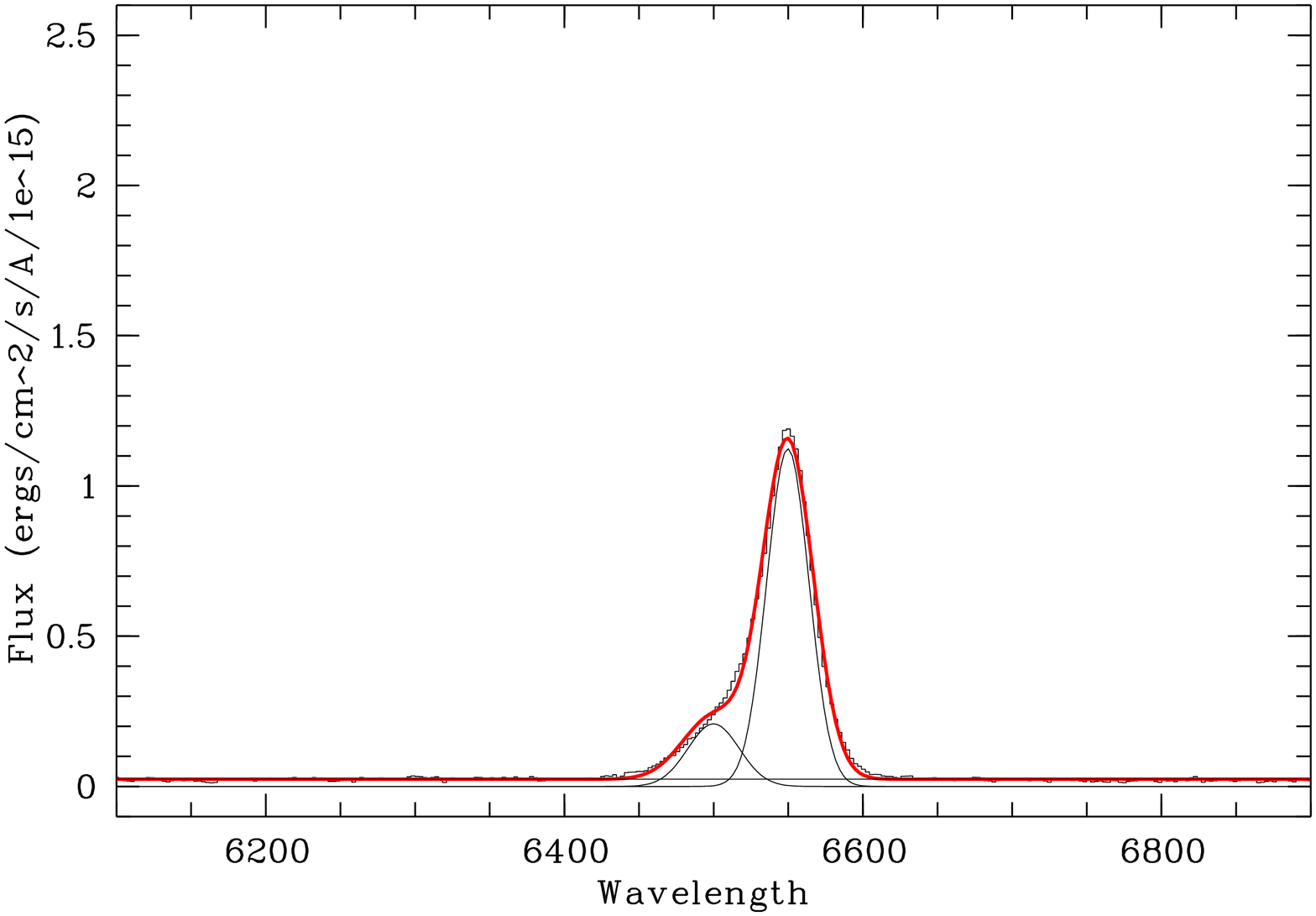, height=90mm, width=60mm , angle=-90}
\caption[]{H$_{\alpha}$ profile from each epoch (102, 131, 160, 208, 240, 273, 475, and 497 days after explosion going left-right from top to bottom). The profiles are decomposed into 2 or more
components representing the broad, intermediate and narrow velocity components.
Note that in the later spectra only the broad and intermediate components are
observed. The parameters of these fits are presented in Table~\ref{hafittbl}.}
\label{hafitonline1} 
\end{figure*}

\begin{figure*}
\epsfig{file=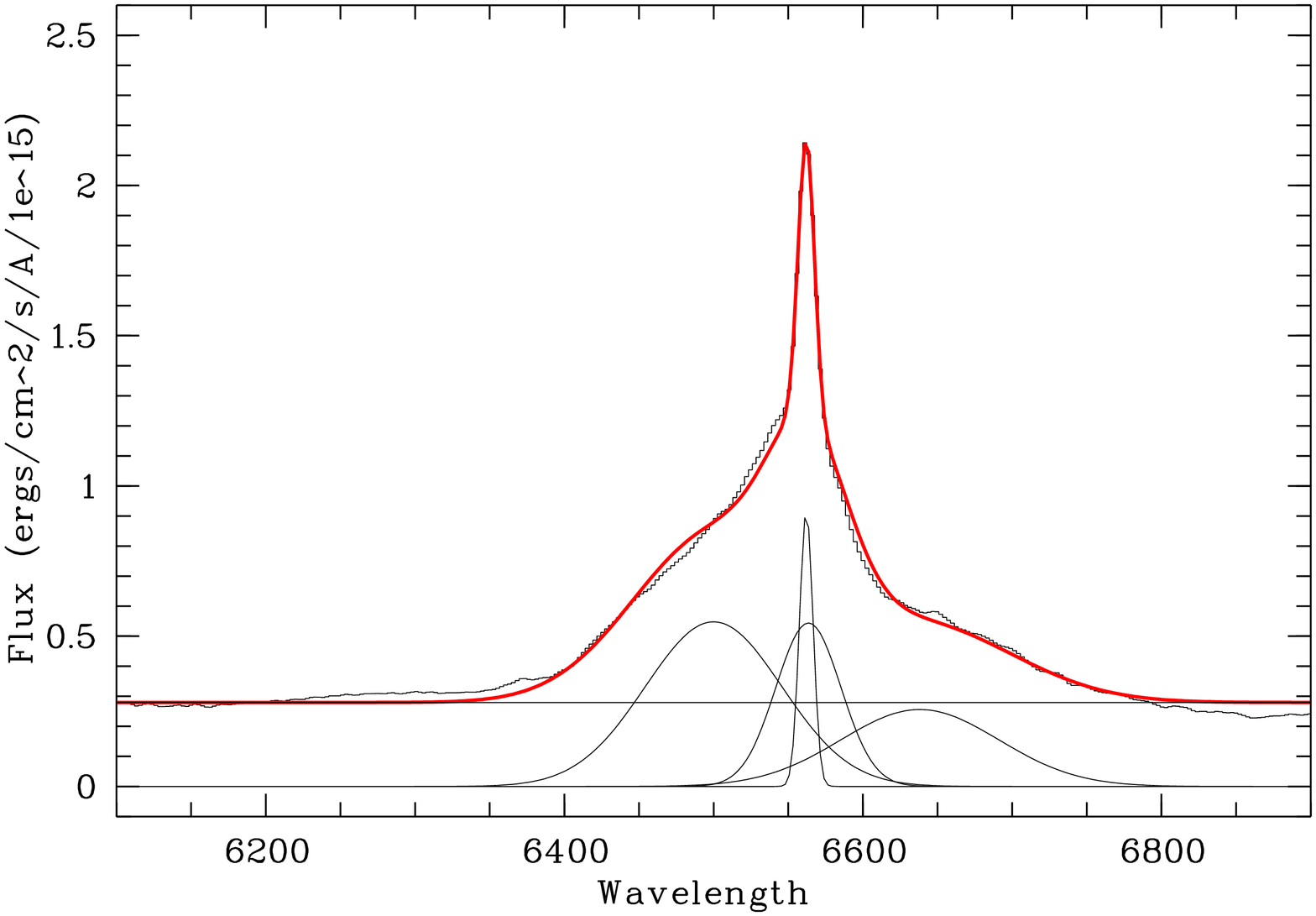, height=90mm, width=62mm , angle=-90}
\epsfig{file=./figures/ha_20080113_blob.ps, height=90mm, width=62mm , angle=-90}
\epsfig{file=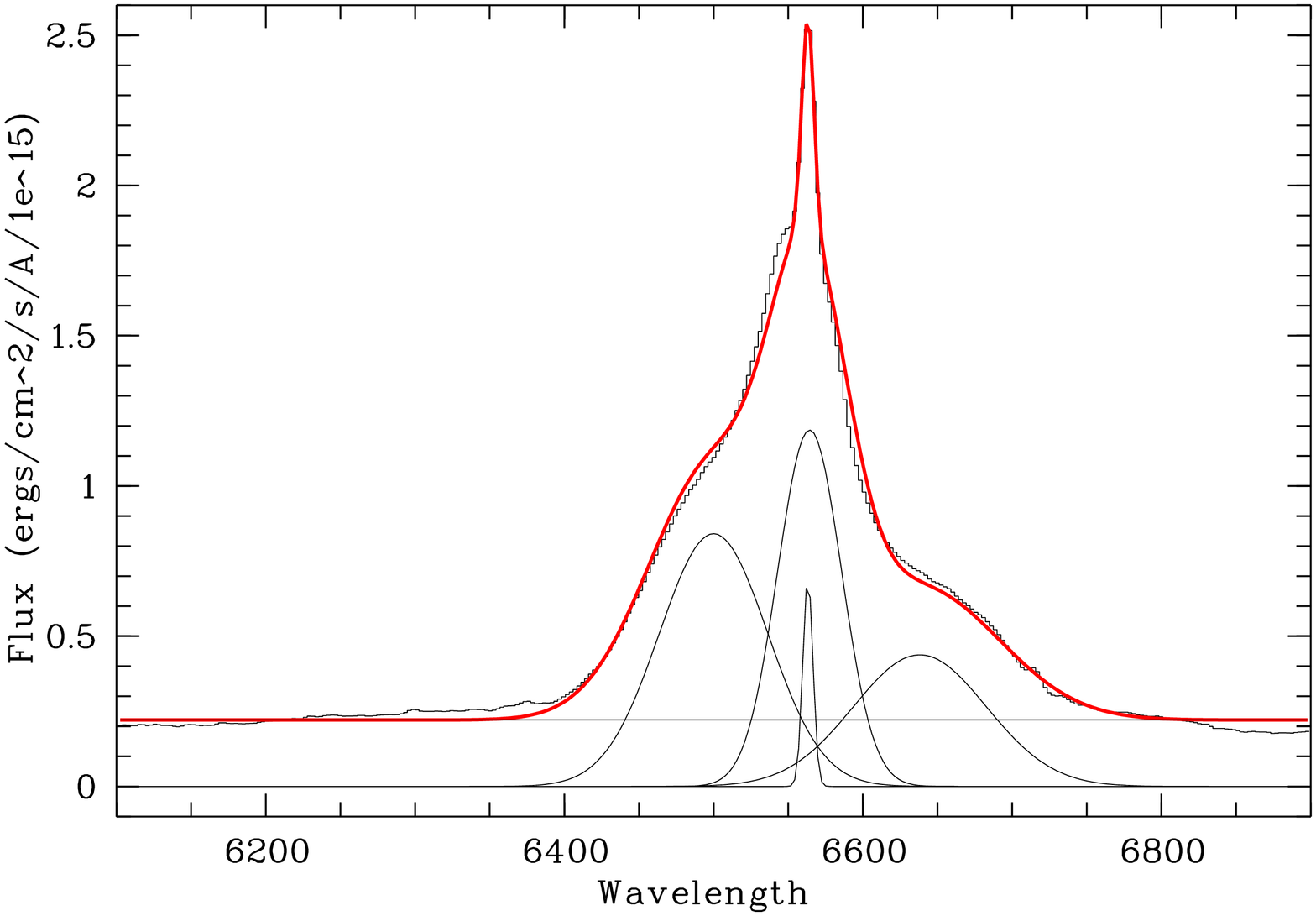, height=90mm, width=62mm , angle=-90}
\epsfig{file=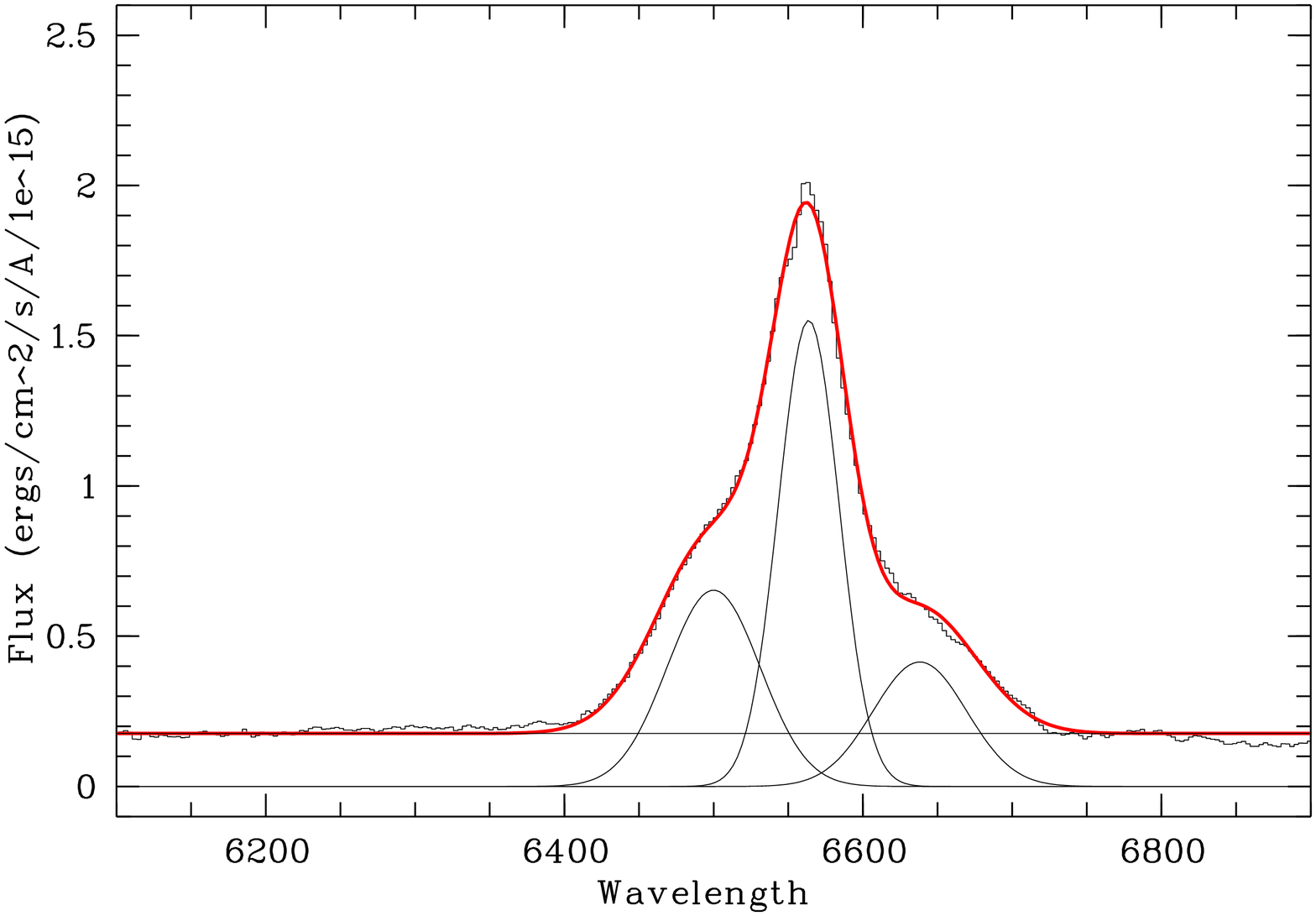, height=90mm, width=62mm , angle=-90}
\epsfig{file=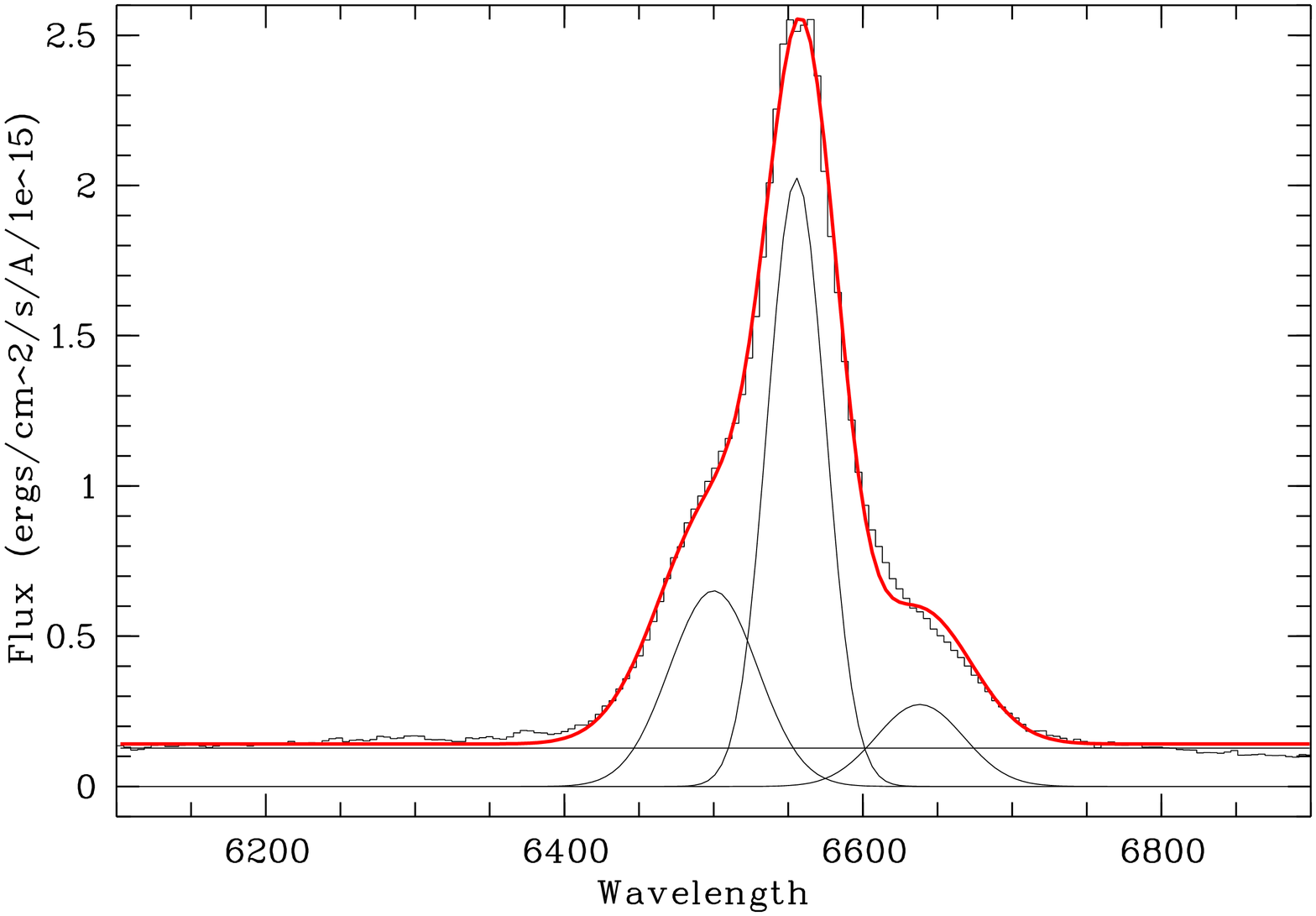, height=90mm, width=60mm , angle=-90}
\epsfig{file=./figures/ha_20081222_blob.ps, height=90mm, width=60mm , angle=-90}
\epsfig{file=./figures/ha_20090114_blob.ps, height=90mm, width=60mm , angle=-90}
\hspace{3mm}\epsfig{file=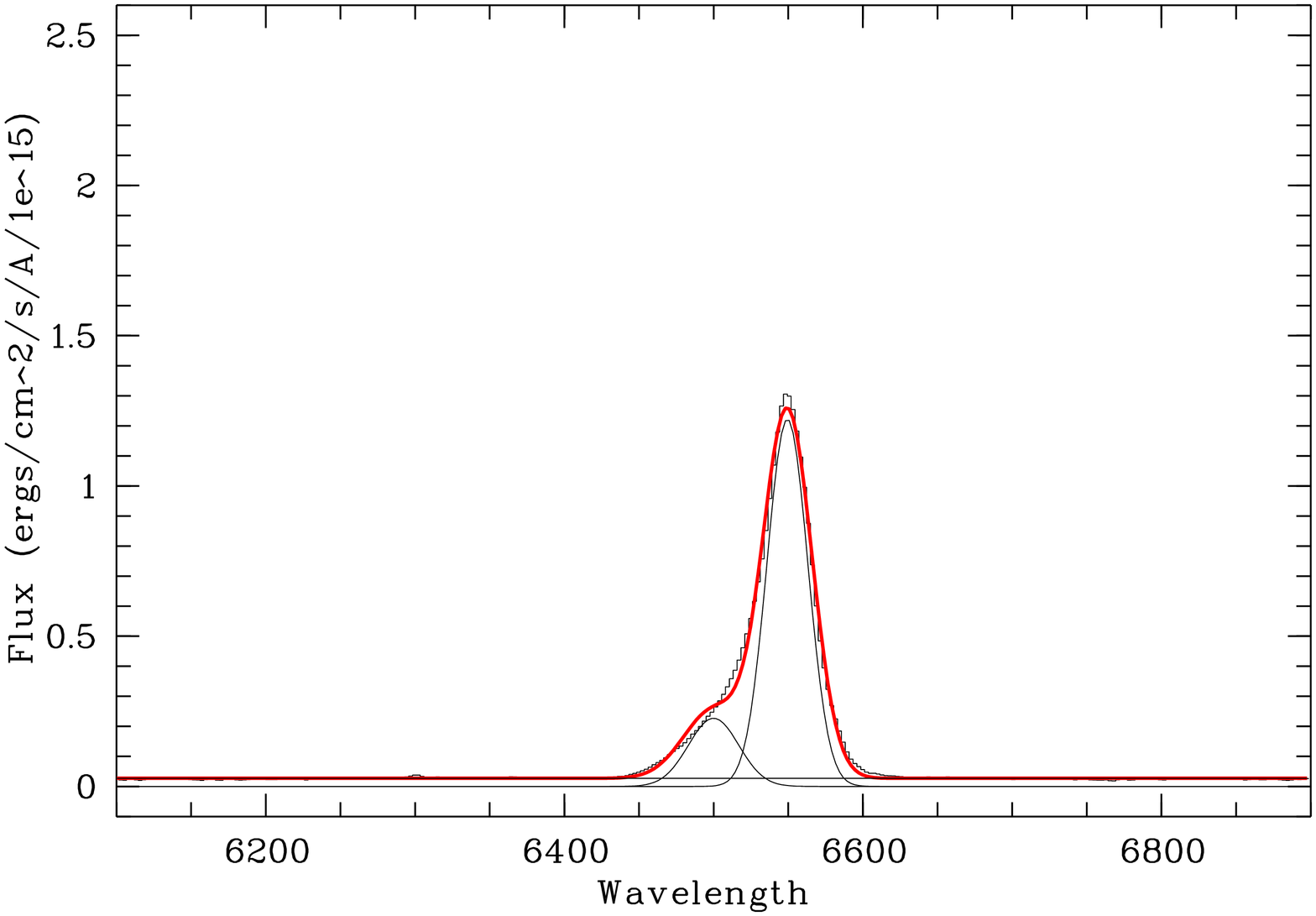, height=90mm, width=60mm , angle=-90}
\caption[]{H$_{\alpha}$ profile from each epoch (102, 131,160, 208, 240, 475, 497, and 507 days after explosion going left-right from top to bottom) decomposed into 3 or more
components representing the two broad, the intermediate and narrow velocity components.
In the later spectra, only the broad and intermediate components are
observed. Parameters for these fits are presented in Table~\ref{hafittbl2}.}
\label{hafitonline2} 
\end{figure*}

\begin{figure*}
\ContinuedFloat
\epsfig{file=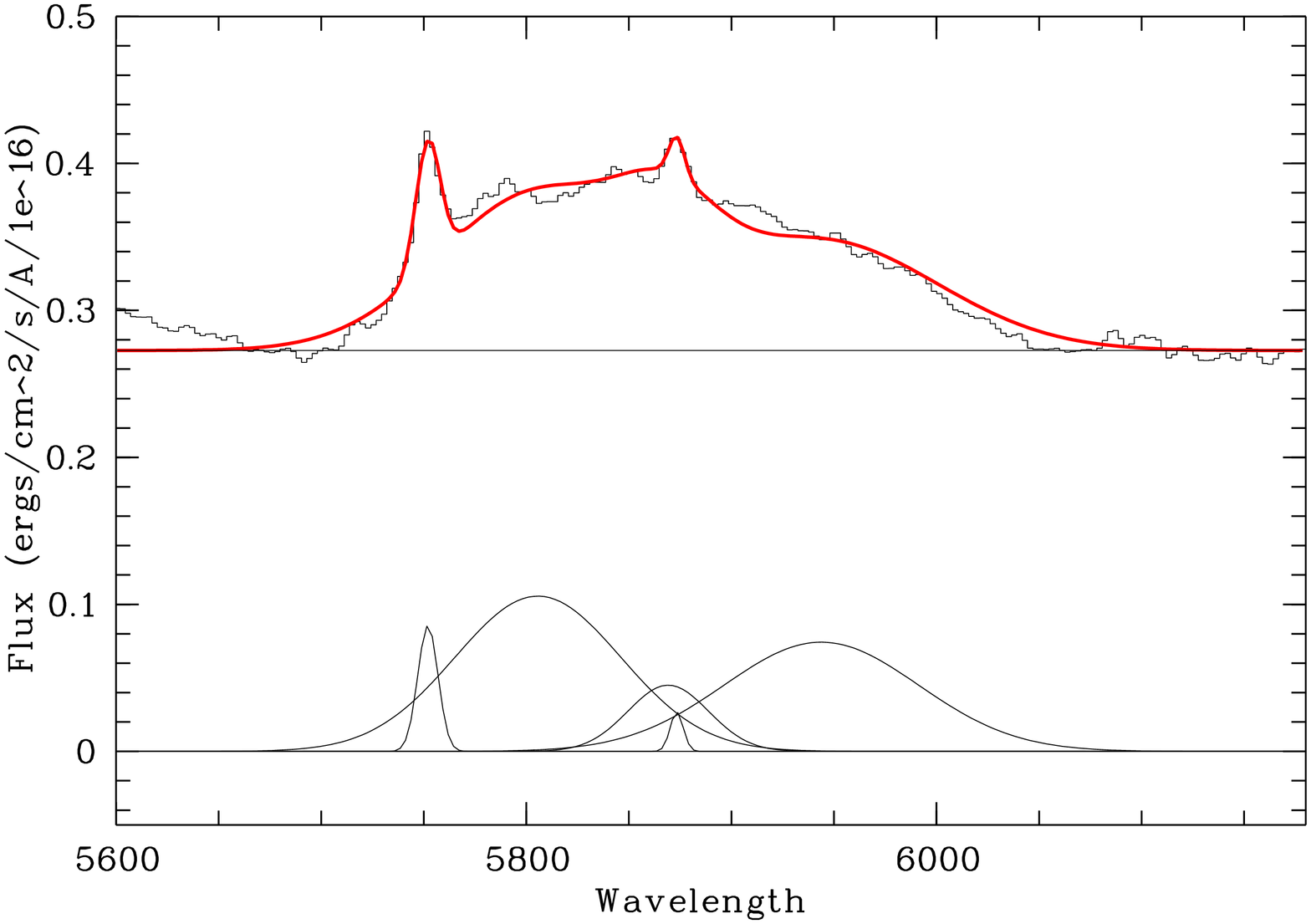, height=90mm, width=62mm , angle=-90}
\epsfig{file=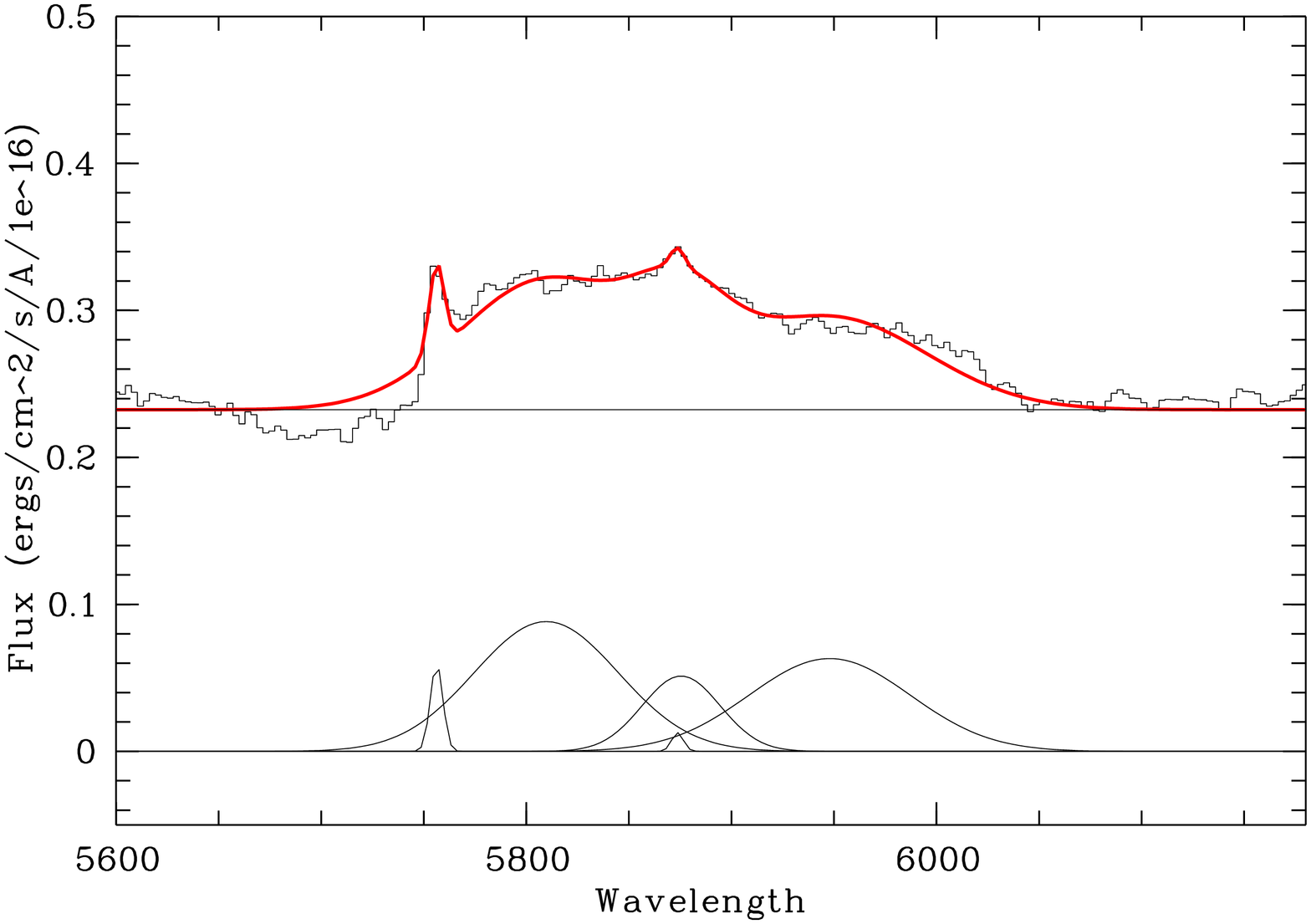, height=90mm, width=62mm , angle=-90}
\epsfig{file=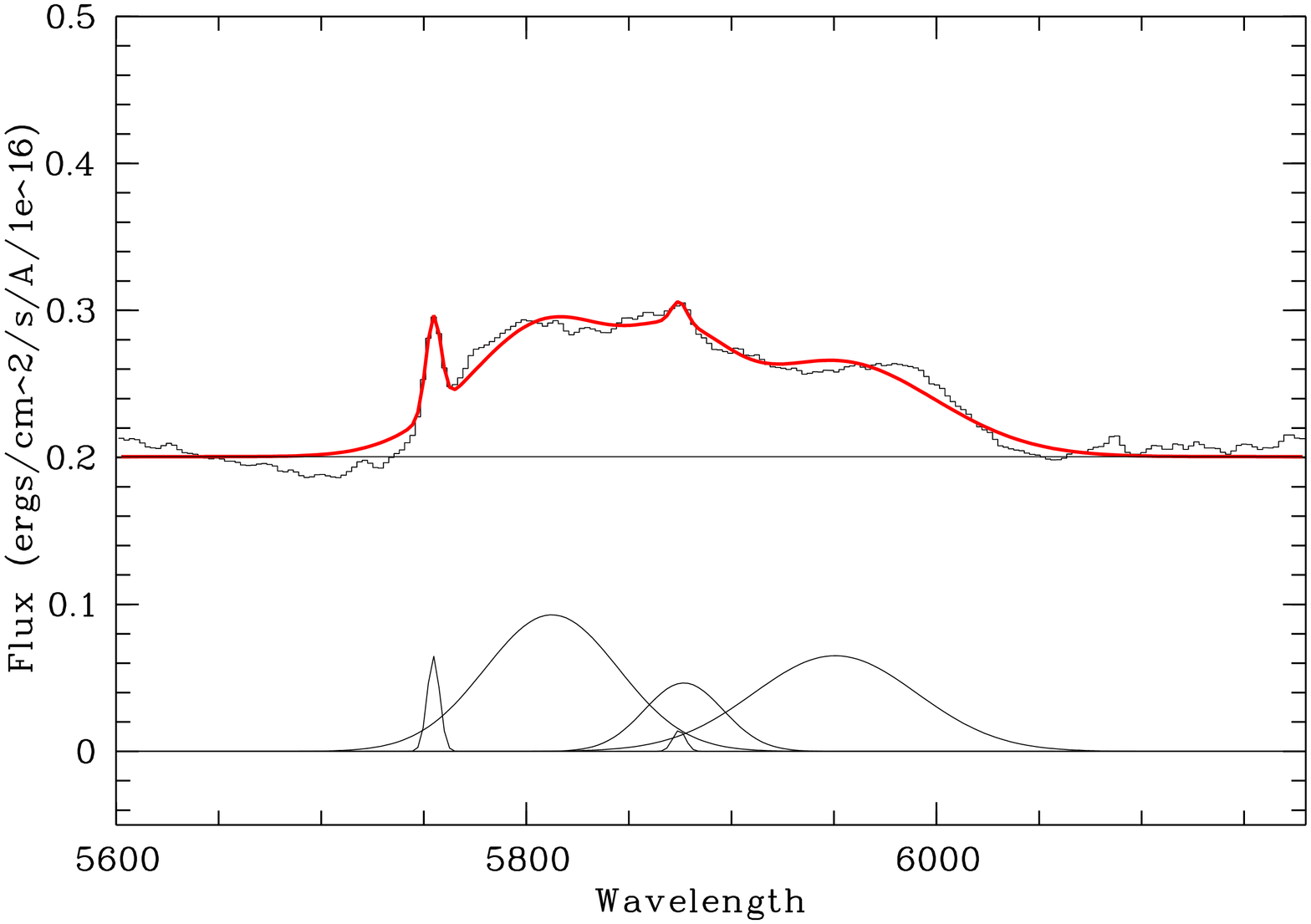, height=90mm, width=62mm , angle=-90}
\epsfig{file=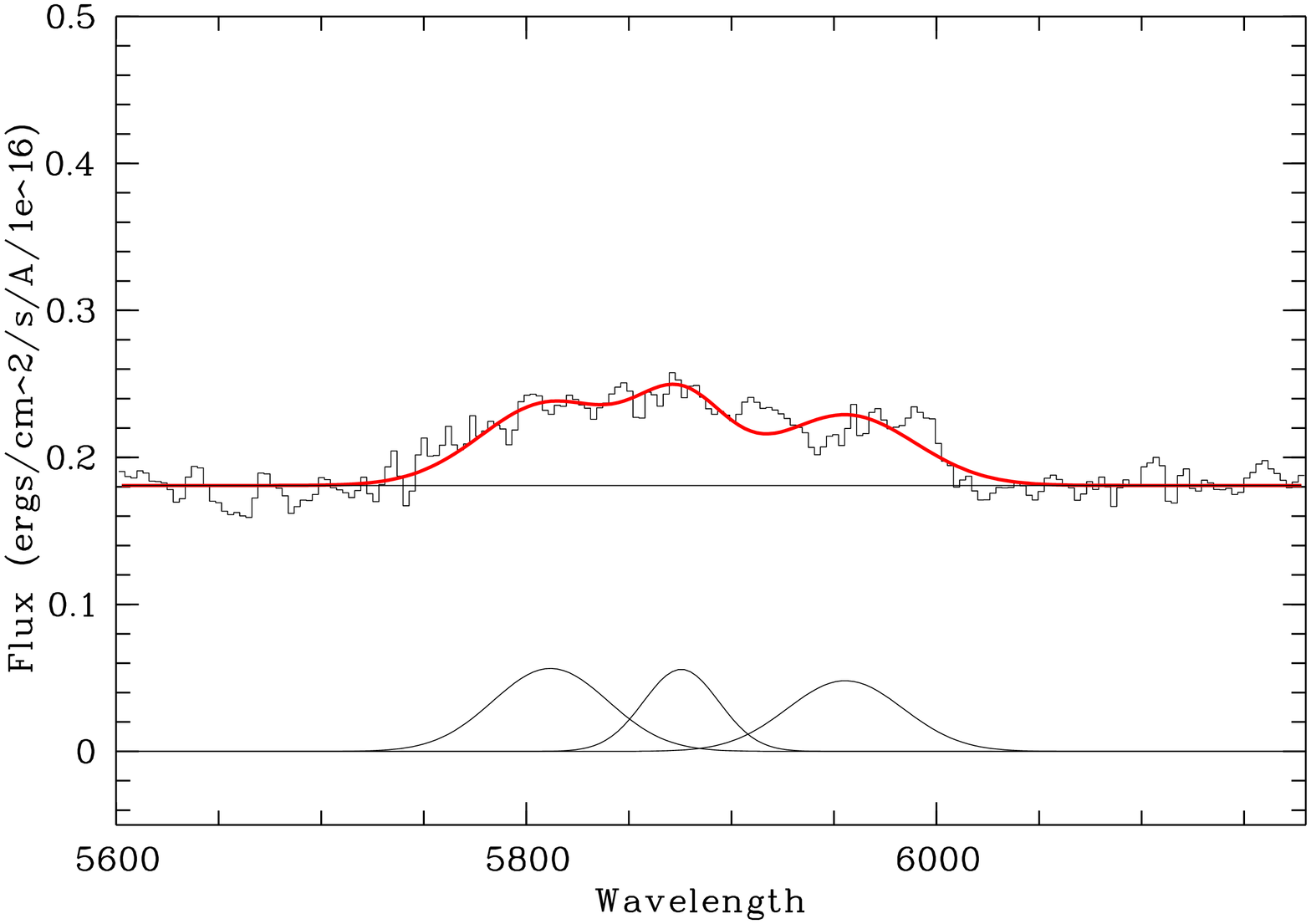, height=90mm, width=62mm , angle=-90}
\epsfig{file=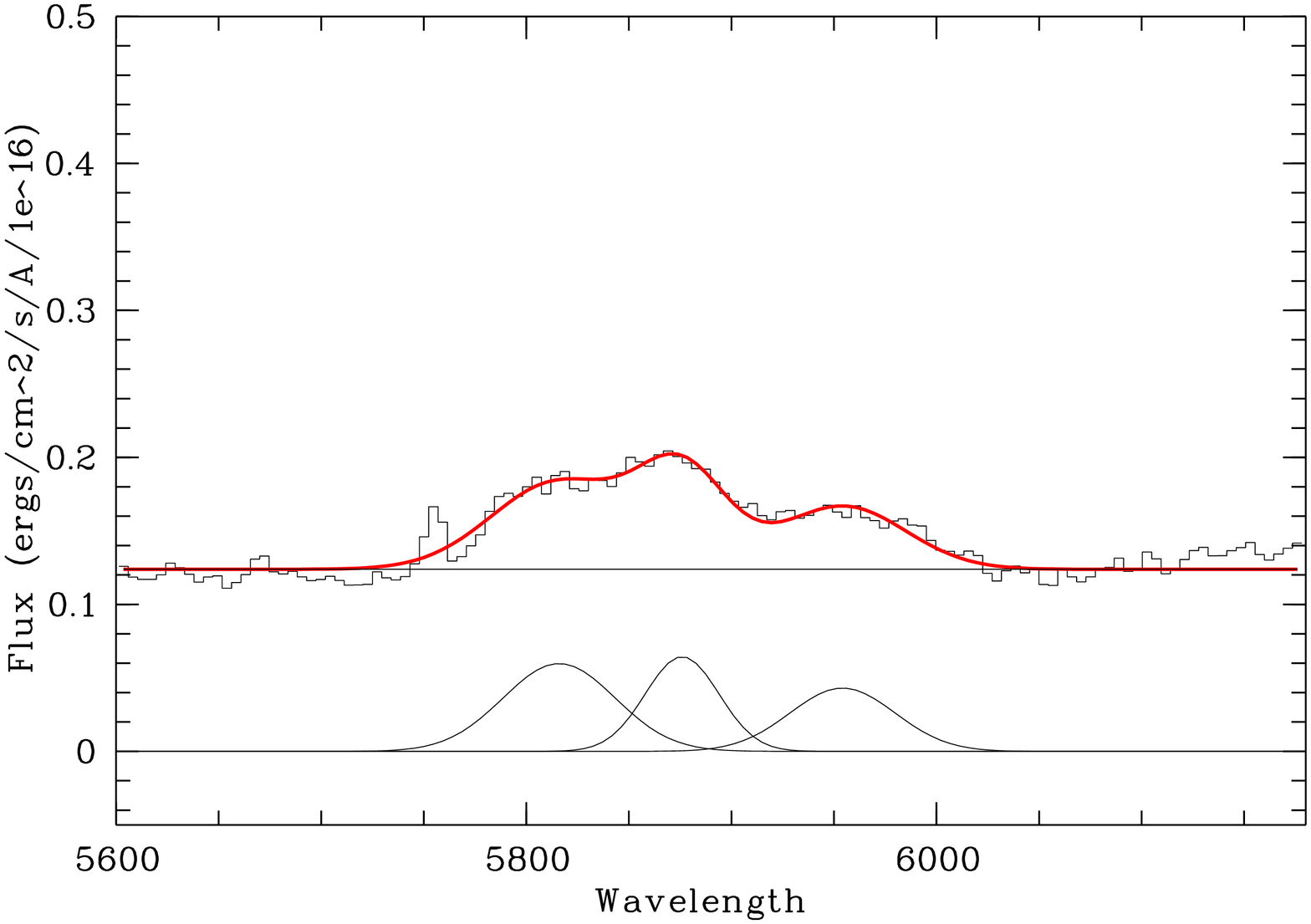, height=90mm, width=62mm , angle=-90}
\epsfig{file=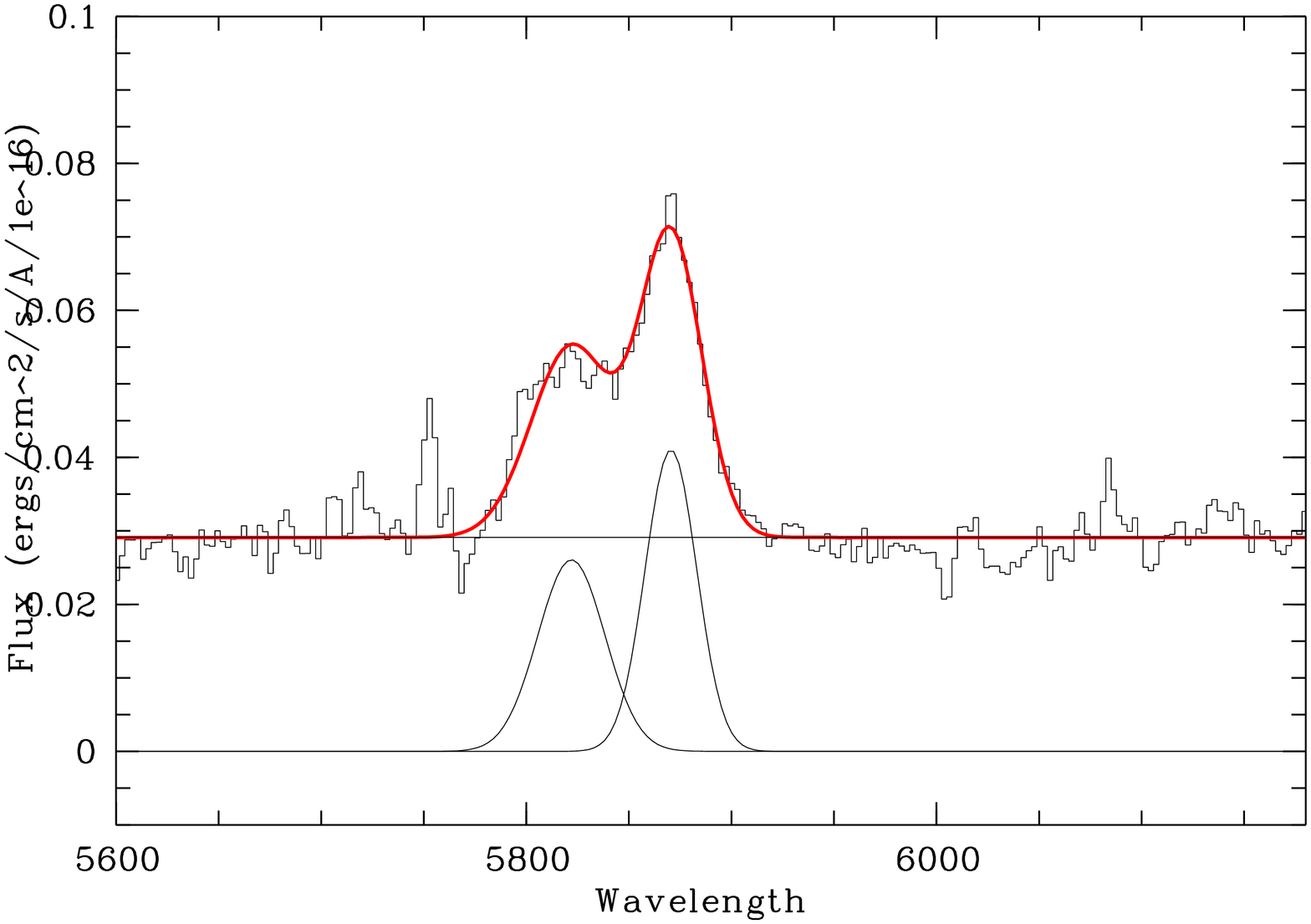, height=90mm, width=62mm , angle=-90}
\epsfig{file=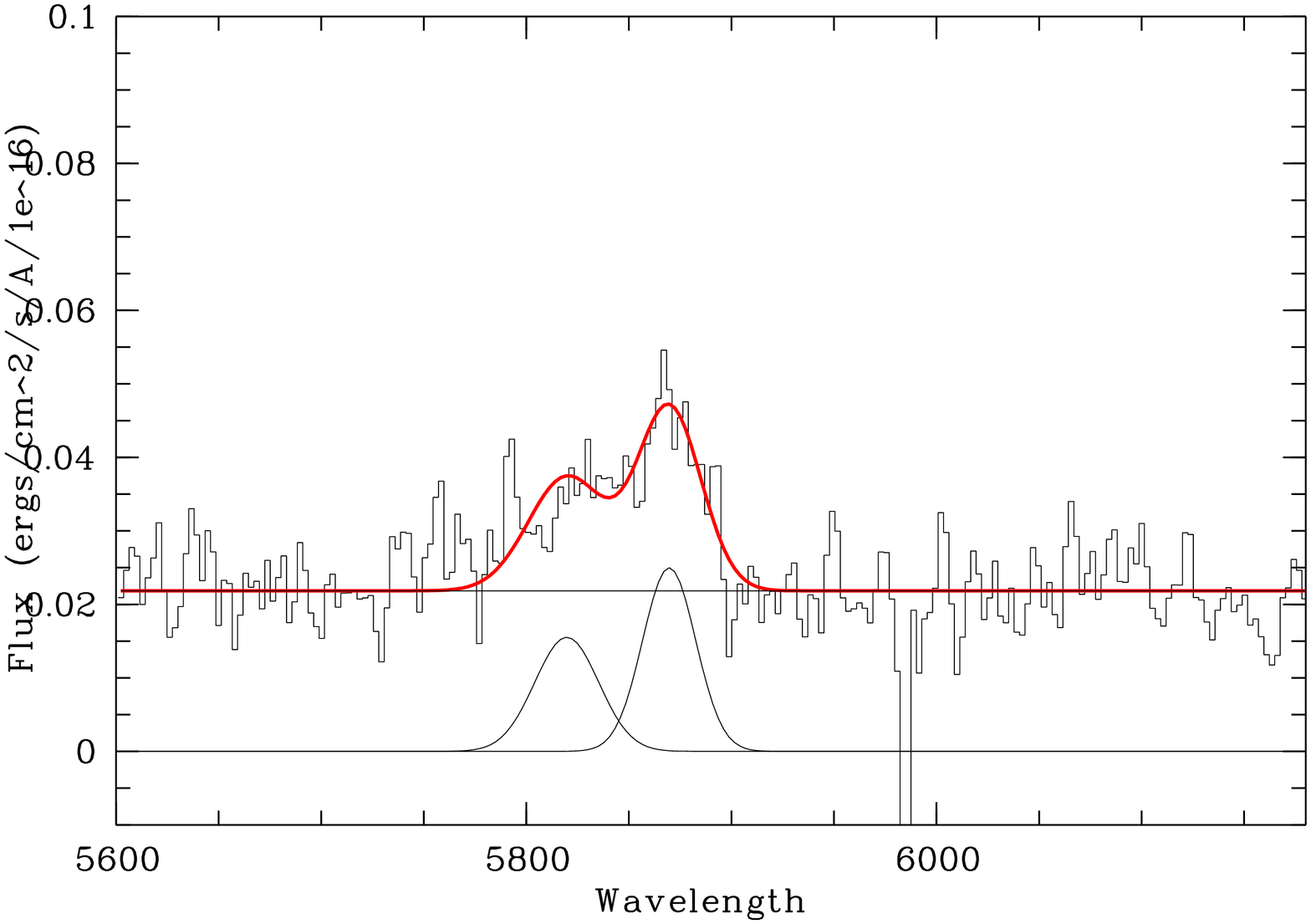, height=90mm, width=62mm , angle=-90}
\hspace{3mm}\epsfig{file=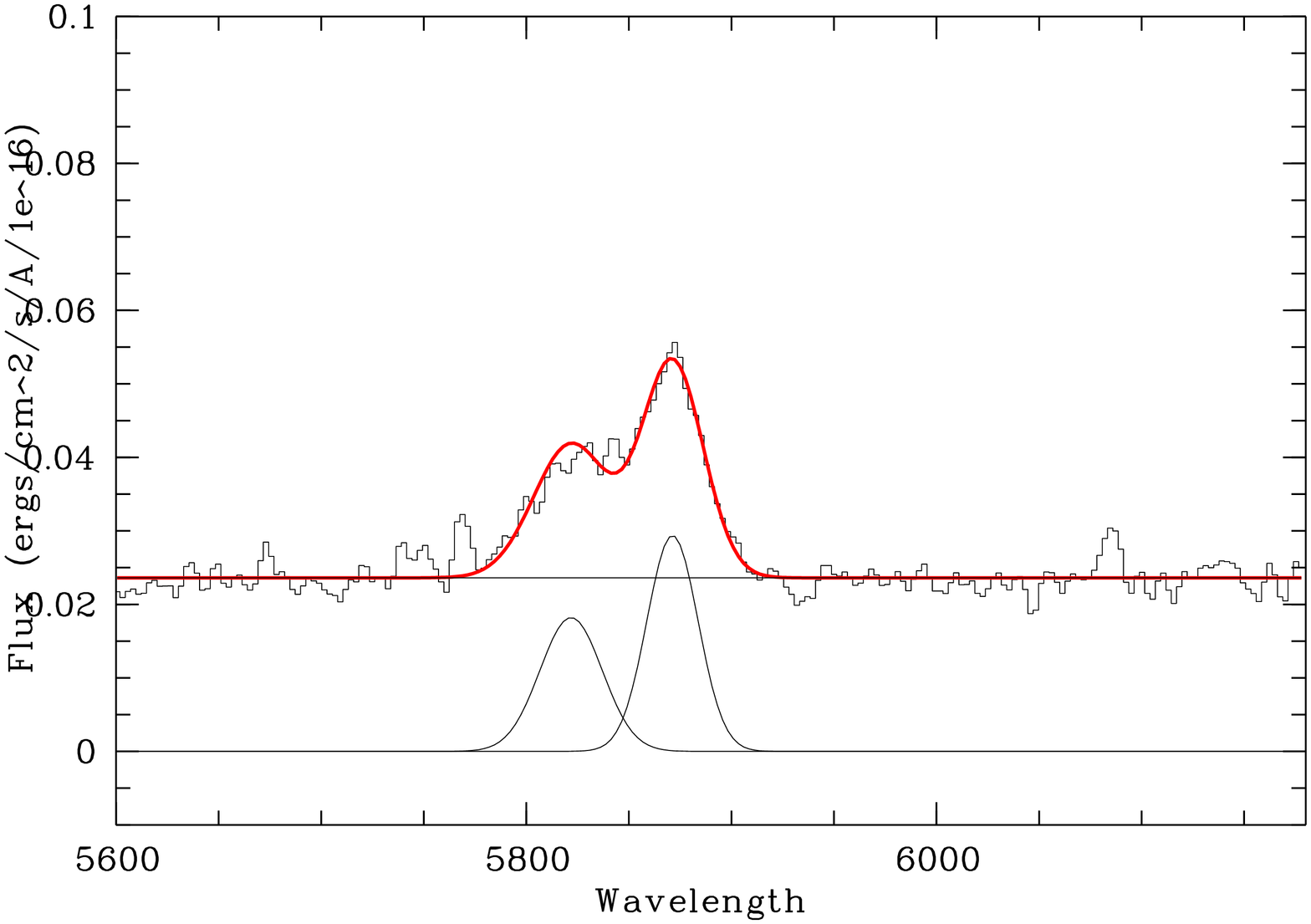, height=90mm, width=62mm , angle=-90}
\caption[]{He {\sc i} 5875 \AA\ profile from each epoch (102, 131,160, 208, 240, 475, 497, and 507 days after explosion going left-right from top to bottom) decomposed into 3 or more
components representing the two broad, the intermediate and narrow velocity components.
In the later spectra, only the broad and intermediate components are
observed. Parameters for these fits are presented in Table~\ref{hafittbl2}.}
\label{hafitonline2contd} 
\end{figure*}

\end{document}